\documentclass[twocolumn, twocolappendix]{aastex631}
\usepackage{physics}
\usepackage{enumitem}

\allowdisplaybreaks[1]
\shorttitle{CO Rovib. Absorption in IRAS 08572+3915 NW}
\shortauthors{Onishi et al.}

\begin{document}

\title{Study of the Inner Structure of the Molecular Torus in IRAS 08572+3915 NW\\with Velocity Decomposition of CO Rovibrational Absorption Lines
\footnote{This research is based on data collected at Subaru Telescope, which is operated by the National Astronomical Observatory of Japan. We are honored and grateful for the opportunity of observing the universe from Maunakea, which has the cultural, historical, and natural significance in Hawaii.}}

\AuthorCollaborationLimit=10
\correspondingauthor{Shusuke Onishi}
\email{s\_onishi@ir.isas.jaxa.jp}

\author[0000-0002-1765-7012]{Shusuke Onishi}
\affiliation{Institute of Space and Astronautical Science, Japan Aerospace Exploration Agency, 3-1-1 Yoshinodai, Chuo-ku, Sagamihara, Kanagawa 252-5210, Japan}
\affiliation{Department of Physics, Graduate School of Science, The University of Tokyo,
7-3-1 Hongo, Bunkyo-ku, Tokyo 113-0033, Japan}


\author[0000-0002-6660-9375]{Takao Nakagawa}
\affiliation{Institute of Space and Astronautical Science, Japan Aerospace Exploration Agency, 3-1-1 Yoshinodai, Chuo-ku, Sagamihara, Kanagawa 252-5210, Japan}

\author[0000-0002-9850-6290]{Shunsuke Baba}
\affiliation{National Astronomical Observatory of Japan, 2-21-1 Osawa, Mitaka, Tokyo 181-8588, Japan}


\author[0000-0002-5012-6707]{Kosei Matsumoto}
\affiliation{Institute of Space and Astronautical Science, Japan Aerospace Exploration Agency, 3-1-1 Yoshinodai, Chuo-ku, Sagamihara, Kanagawa 252-5210, Japan}
\affiliation{Department of Physics, Graduate School of Science, The University of Tokyo,
7-3-1 Hongo, Bunkyo-ku, Tokyo 113-0033, Japan}

\author{Naoki Isobe}
\affiliation{Institute of Space and Astronautical Science, Japan Aerospace Exploration Agency, 3-1-1 Yoshinodai, Chuo-ku, Sagamihara, Kanagawa 252-5210, Japan}

\author{Mai Shirahata}
\affiliation{Institute of Space and Astronautical Science, Japan Aerospace Exploration Agency, 3-1-1 Yoshinodai, Chuo-ku, Sagamihara, Kanagawa 252-5210, Japan}

\author[0000-0002-7914-6779]{Hiroshi Terada}
\affiliation{National Astronomical Observatory of Japan, 2-21-1 Osawa, Mitaka, Tokyo 181-8588, Japan}

\author[0000-0001-9855-0163]{Tomonori Usuda}
\affiliation{National Astronomical Observatory of Japan, 2-21-1 Osawa, Mitaka, Tokyo 181-8588, Japan}

\author[0000-0003-4842-565X]{Shinki Oyabu}
\affiliation{Institute of Liberal Arts and Sciences, Tokushima University, 1-1 Minami-josanjima-cho, Tokushima, Tokushima 770-8502, Japan}




\begin{abstract}
Understanding the inner structure of the clumpy molecular torus surrounding the active galactic nucleus is essential in revealing the forming mechanism. However, spatially resolving the torus is difficult because of its size of a few parsecs. Thus, to probe the clump conditions in the torus, we performed the velocity decomposition of the CO rovibrational absorption lines ($\Delta{v}=0\to 1,\ \Delta{J}=\pm 1$) at $\lambda\sim 4.67\,\mathrm{\mu{m}}$ observed toward an ultraluminous infrared galaxy IRAS 08572+3915 NW with the high-resolution spectroscopy ($R\sim 10{,}000$) of Subaru Telescope. Consequently, we found that each transition had two outflowing components, i.e., (a) and (b), both at approximately $\sim -160\,\mathrm{km\,s^{-1}}$, but with broad and narrow widths, and an inflowing component, i.e., (c), at approximately $\sim +100\,\mathrm{km\,s^{-1}}$, which were attributed to the torus. The ratios of the velocity dispersions of each component lead to those of the rotating radii around the black hole of $R_\mathrm{rot,a}:R_\mathrm{rot,b}:R_\mathrm{rot,c}\approx 1:5:17$, indicating the torus where clumps are outflowing in the inner regions and inflowing in the outer regions if a hydrostatic disk with $\sigma_V\propto R_\mathrm{rot}^{-0.5}$ is assumed. Based on the kinetic temperature of components (a) and (b) of $\sim 720\,\mathrm{K}$ and $\sim 25\,\mathrm{K}$ estimated from the level population, the temperature gradient is $T_\mathrm{kin}\propto R_\mathrm{rot}^{-2.1}$. Magnetohydrodynamic models with large density fluctuations of two orders of magnitude or more are necessary to reproduce this gradient.
\end{abstract}



\section{Introduction}\label{sec:intro}
Active galactic nuclei (AGNs) are classified into type 1 and type 2 based on their optical line width. The unified model of AGNs \citep[e.g.,][]{Miller1983,Antonucci1985a,Antonucci1993b} suggests that the inclination of some geometrically thick structure surrounding the central black hole, which is the molecular torus, predominantly makes the difference. Then, in order to clarify the mechanism maintaining the geometrical thickness, it is essential to understand its inner structure. However, spatially resolving the torus is difficult because its size is expected to be a few parsecs. Although the radio interferometry or polarimetry with torus-scale beams is being achieved toward the nearest AGNs, such as NGC 1068 \citep[e.g.,][]{Garcia-Burillo2016,Imanishi2018,Imanishi2020, Lopez-Rodriguez2020}, these days, the structure therein has not yet been resolved.\par
For the above reason, the inner structure of the torus is mainly discussed by theoretical models. It is expected that the torus consists of many dense molecular clouds (clumps), and the turbulence or outflow motion works better to maintain its geometrical thickness in the clump--clump collisional disk than in the continuous gas disk (clumpy torus models; \citealp[e.g.,][]{Beckert2004,Vollmer2004,Nenkova2008a}). The clumpy torus models are also required to reproduce the observed optical depth of silicate dust at $\lambda\sim 9.7\,\mathrm{\mu{m}}$ \citep[e.g.,][]{Nenkova2002a, Dullemond2005a}. Here, we should note that not all the mid-infrared flux at the wavelength of $\lambda\sim 8\text{--}13\,\mathrm{\mu{m}}$ originates in the torus; some of it comes from the polar dust, according to the recent mid-infrared interferometry \citep[e.g.,][]{Honig2013,Tristram2014,Asmus2019}.\par
\begin{figure}
    \centering
    \includegraphics[width=\linewidth]{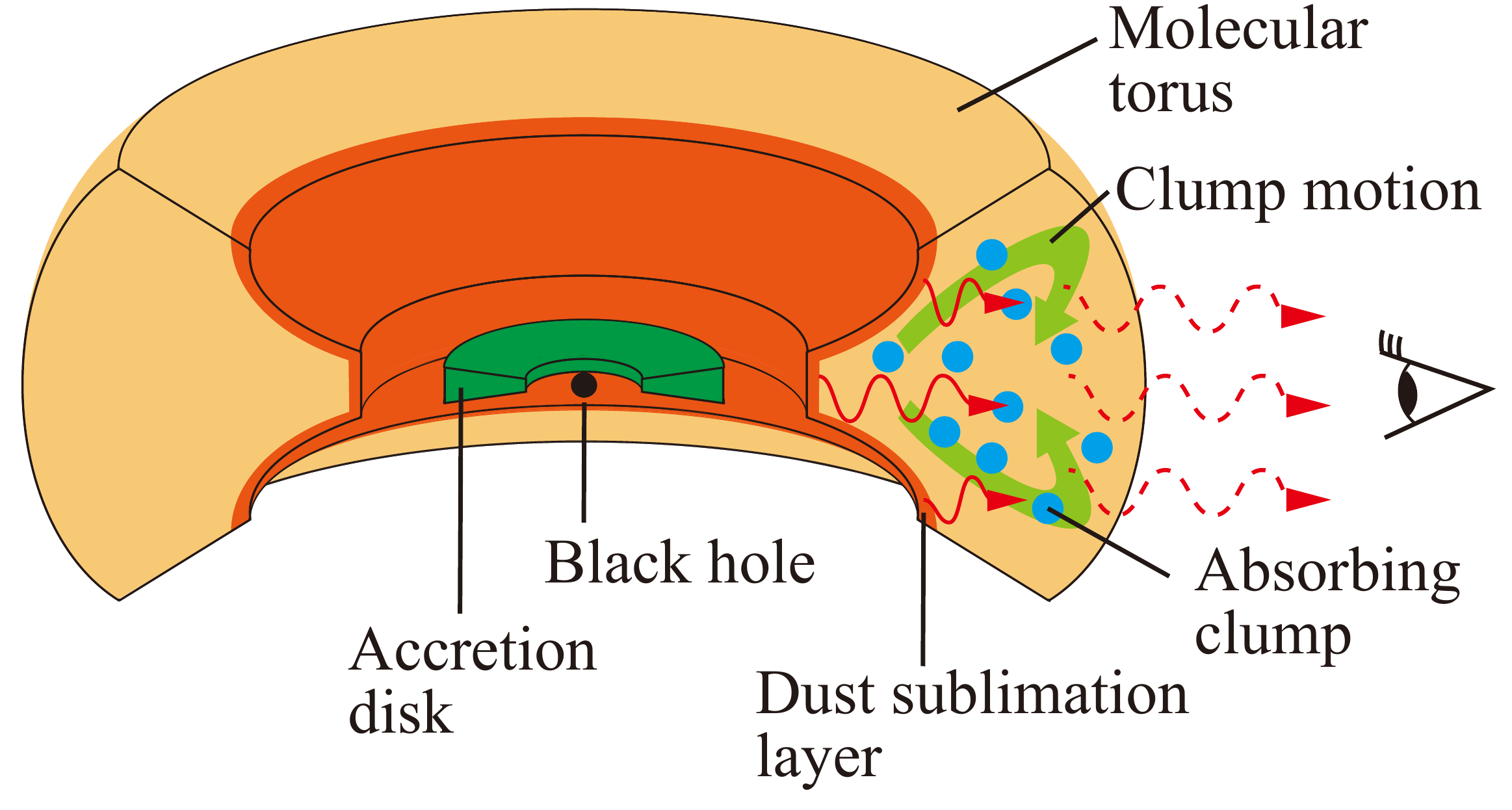}
    \caption{Simplified schematic image of the geometry of the dust sublimation layer (the continuum source) and the absorbing clumps that cause CO rovibrational absorption lines.}
    \label{fig:torus_clumpy}
\end{figure}

\citet{Wada2012a} and \citet{Wada2016} proposed the  ``radiation fountain model,'' where the molecular torus was formed by the outflowing and inflowing gas around the black hole and the accretion disk. This is also supported by some other radiation hydrodynamic (RHD) and magnetohydrodynamic (MHD) simulations \citep[e.g.,][]{Namekata2016,Chan2017,Kudoh2020,Venanzi2020}. The molecular torus is then predicted not as a static structure but as a dynamic one. Thus, the clumps in the torus are expected to be inflowing or outflowing as shown in the simplified schematic image in Figure \ref{fig:torus_clumpy}.
Observational studies on the dynamical and physical properties and the spatial distributions of the clumps are required to clarify the inner structure of the torus.\par
To estimate the clump conditions, we observe herein the absorption band of CO rovibrational transitions ($v=0\to 1,\ \Delta{J}=\pm{1}$) at $\lambda\sim 4.67\,\mathrm{\mu{m}}$. The observation of the CO rovibrational absorption band has two advantages. First, we can preferentially observe the CO absorption by the clumps in the torus, avoiding the host galaxy contamination because the main near-infrared (NIR) continuum source at $\lambda\sim 4.67\,\mathrm{\mu{m}}$ is expected to be hot dust at the dust sublimation layer whose radius is $\sim 1\,\mathrm{pc}$ \citep{Rees1969a,Rieke1978,Barvainis1987,GRAVITYCollaboration2020a}. Figure \ref{fig:torus_clumpy} illustrates the assumed geometry of the dust sublimation layer and the clumps. Second, we can simultaneously observe many absorption lines with multiple rotational levels and determine the level population at $v=0$ from their optical depth. This leads to an accurate estimate of the physical properties, such as the excitation temperature and the molecular column density.\par
In this way, the previous low-resolution spectroscopy of the CO rovibrational absorption band supports the assumption that it is mainly caused by the warm gas of AGNs, although each transition was not resolved. \citet{Spoon2004} found that the absorbing CO gas was warm ($T\sim 720\,\mathrm{K}$) in an ultraluminous infrared galaxy (ULIRG) IRAS 00183$-$7111 by comparing the observed low-resolution spectrum of the CO rovibrational absorption band derived with Spitzer ($R\sim 80$) and a local thermodynamic equilibrium (LTE) and isothermal slab model \citep{Cami2002}. Moreover, \citet{Baba2018} found that the absorbing CO gas had a warmer excitation temperature in 10 nearby ULIRGs ($T\sim 200\text{--}500\,\mathrm{K}$) than in a typical starburst ($T\lesssim 100\,\mathrm{K}$) based on the model of \citet{Cami2002} using low-resolution spectra of AKARI ($R\sim 150$) and Spitzer ($R\sim 80$).\par
Thus, if we resolve the velocity components outflowing or inflowing in each transition with the high-resolution spectroscopy, we obtain information not only on the physical properties of clumps of each velocity component but also on the dynamical properties of the clumps, such as the line-of-sight (LOS) velocity and the velocity dispersion. The relative spatial distributions of the inflowing and outflowing clumps can then be estimated from the velocity dispersion because it gets larger near the central black hole. For the above reasons, the high-resolution spectroscopy of the CO rovibrational absorption band is a suitable probe into the inner structure of the molecular torus.\par
Actually, \citet{Geballe2006a} and \citet{Shirahata2013b} resolved each velocity component in the CO rovibrational absorption lines with the high-resolution spectroscopy of the United Kingdom Infrared Telescope ($R\sim 7500$) and Subaru Telescope ($R\sim 5000$), respectively, toward the northwest (NW) core of a ULIRG IRAS 08572+3915. \citet{Shirahata2013b} found three velocity components of the relative LOS velocity to the host galaxy of $V-V_\mathrm{sys}\sim -160$ (outflowing), 0 (systemic), and $+100\,\mathrm{km\,s^{-1}}$ (inflowing). They also found that the equivalent widths of each line were reproduced by the sum of two Boltzmann distributions, whose excitation temperatures were $T\sim 25$ and $270\,\mathrm{K}$ under the optically thin assumption, attributing the two temperatures to the systemic and outflowing components, respectively.\par
However, where in the molecular torus each velocity component is caused remains unclear because they did not perform any velocity decomposition to determine the accurate properties of each component. Hence, we performed herein the velocity decomposition of the CO rovibrational absorption lines in the ULIRG IRAS 08572+3915 NW with a higher-resolution spectrum ($R\sim 10{,}000$) than the previous studies to probe the dynamical and physical properties and the spatial distribution of each component.\par
We report the results in this paper. This is the first velocity decomposition study of the CO rovibrational absorption lines in AGNs. Section \ref{sec:target} describes the physical conditions of the target, IRAS 08572+3915 NW. Section \ref{sec:obs_data} presents the observational conditions and the data reduction method. Section \ref{sec:analysis} explains the continuum placement, the subtraction of extra line features and bad data points, and the velocity decomposition. Subsequently, Section \ref{sec:origin_comp} shares the estimated dynamical and physical properties and spatial distributions of each velocity-decomposed clump. Finally, Section \ref{sec:discuss} presents a comparison of the derived properties with some theoretical models of the molecular torus, and Section \ref{sec:conclusion} gives the conclusion.

\section{Target}\label{sec:target}
The observation target in this work, which is IRAS 08572+3915 (hereafter referred to as IRAS08), is a ULIRG whose infrared luminosity is $\log[L(8\text{--}1000\,\mathrm{\mu{m}})/L_\odot]=12.09$ \citep{Sanders1988a}. Its redshift determined by the CO ($J=1\to 0$) emission line is $z=0.0583$ \citep{Evans2002}. IRAS08 has two NIR cores with a separation of $5\farcs4=5.6\,\mathrm{kpc}$ \citep{Scoville2000} in the northwest (NW) and southeast (SE), and approximately $80\%$ of mid-infrared luminosity comes from the NW core \citep{Soifer2000}.\par
In this work, we focus on the NW core. IRAS08 NW is classified as an AGN based on the lack of the polycyclic aromatic hydrocarbon (PAH) emission feature at $\lambda\sim 3.3\,\mathrm{\mu{m}}$, and it is likely to be heavily dust obscured based on the absorption feature of the carbonaceous dust at $\lambda\sim 3.4\,\mathrm{\mu{m}}$ \citep{Imanishi2000,Doi2019} and the silicate dust at $\lambda\sim 9.7\,\mathrm{\mu{m}}$ of $\tau_\mathrm{Si,9.7}\sim 5.2$ \citep{Dudley1997}. Moreover, this AGN is expected to be Compton thick ($N_\mathrm{H}\gg 10^{24}\,\mathrm{cm^{-2}}$) based on a small number of hard X-ray counts ($\sim 10$) and a very small ratio of the hard X-ray luminosity to the infrared luminosity \citep{Iwasawa2011}. \citet{Efstathiou2014} estimated the intrinsic AGN luminosity of IRAS08 NW as $L_\mathrm{AGN}\approx 9\times 10^{45}\,\mathrm{erg\,s^{-1}}$ by the model fitting to its spectral energy distribution (SED) from the infrared to submillimeter wavelength correcting the emission anisotropy of the molecular torus. The radio interferometry toward IRAS08 NW has not spatially resolved its molecular torus because the highest spatial resolution ever achieved is $0\farcs18=187\,\mathrm{pc}$ \citep{Imanishi2018a} by Atacama Large Millimeter/submillimeter Array (ALMA).

\section{Observation and Data Reduction}\label{sec:obs_data}
\subsection{Subaru IRCS Observation}
We conducted $M$-band echelle spectroscopy toward IRAS08 NW with the Infrared Camera and Spectrograph (IRCS; \citealp{Tokunaga1998,Kobayashi2000}) of the $8.2\,\mathrm{m}$ Subaru Telescope \citep{Iye2004} on Maunakea, Hawaii, for three nights in 2010 and 2019. Table \ref{tab:obsinfo} summarizes the observational information of the spectral data used herein. In all observations, $0\farcs27\times 9\farcs37$ apertures were used, corresponding to the spectral resolution of $R\sim 10{,}000$ or the velocity resolution of $\Delta{V}\sim 30\,\mathrm{km\,s^{-1}}$. The slit position angles were set to $\text{PA}=55^\circ$ east of north to avoid the SE core of IRAS08. These observations covered a wavelength ranging from 4.73 to 5.13 $\mathrm{\mu{m}}$, including CO rovibrational lines with $0\le J\le 26$ in the $R$-branch ($v=0\to 1,\ J\to J+1$), and $1\le J\le 19$ in the $P$-branch ($v=0\to 1,\ J\to J-1$), at $z=0.0583$. To improve the signal-to-noise ratio (S/N), the adaptive optics, AO188 \citep{Hayano2008,Hayano2010}, with laser guide stars was used for the 2019 observations, but not for those in 2010. Thus, the seeing size in 2010 was larger than that in 2019. For the sky subtraction, all observations were conducted in the A-B-B-A nodding mode, in which the telescope was nodded for 4\farcs0 along the slit.

\begin{deluxetable*}{cccccccc}
\tablecaption{Observation Log for IRAS 08572+3915 NW
\label{tab:obsinfo}}
\tablehead{
\colhead{No.} & \colhead{Obs. ID} & \colhead{Date (UT)} & \colhead{(ECH, XDS)} & \colhead{$\lambda\,\mathrm{(\mu m)}$} & \colhead{Int. Time (minutes)} & \colhead{AO} & \colhead{Seeing (arcsec)}
}
\decimalcolnumbers
\startdata
1 & o10405 & 2010/3/1 & $(-3200, -5500)$ & 4.73--4.84 & 72 & No & 0.8\\
2 & o18163 & 2019/1/19 & $(6500, -6100)$ & 5.00--5.13 & 160 & Yes & 0.3\\
3 & o18163 & 2019/1/20 & $(12000, -5650)$ & 4.90--5.04 & 128 & Yes & 0.4\\
4 & o18163 & 2019/1/20 & $(-10000, -6100)$ & 4.82--4.92 & 100.8 & Yes & 0.4\\
\enddata
\tablecomments{
Column (1): data number in this paper. Column (2): observation ID given by Subaru Telescope. Column (3): observation date. Column (4): unique configuration number for the angles of the echelle grating (ECH) and the cross disperser (XDS) of Subaru IRCS. Column (5): observed wavelength ranges. Column (6): on-source integration time. Column (7): with or without AO. Column (8): FWHM seeing size in $K$ band.
}
\end{deluxetable*}

\subsection{Data Reduction}\label{subsec:data_reduc}
One-dimensional raw spectra were extracted from the slit images using IRAF v2.16.1 \citep{Tody1986,Tody1993} via PyRAF v2.1.15 \citep{Pyraf} in a standard manner. To minimize the systematic error caused by the wavelength calibration, we fit the telluric absorption lines imprinted in the spectra of the standard stars with the telluric line model using Molecfit v1.5.9 packages \citep{Kausch2015,Smette2015}. Table \ref{tab:std_stars} summarizes the standard-star parameters for each data group. After the wavelength calibration, the IRAS08 NW spectra were divided by the standard-star spectra and multiplied by the blackbody spectra with the corresponding effective temperature ($T_\mathrm{eff}$) to correct the features of the telluric atmosphere and the bias of the throughput. To minimize the difference of the airmass between IRAS08 NW and the standard stars, we chose standard stars such that they have an air mass within 0.15 from that of IRAS08 NW. We evaluated the flux error by separating the observational data with each configuration into four groups based on whether the slit is on the A or B position of A-B-B-A nodding and whether the data were derived in the former (f) or latter (l) half of each observational sequence. We then adopted the standard deviation of each flux derived from the four groups (i.e., A-f, A-l, B-f, and B-l) as the flux error, assuming Student's $t$-distribution.

\begin{deluxetable}{ccccccc}
\tablecaption{Data Groups and Their Standard Stars
\label{tab:std_stars}}
\tablehead{
\colhead{No.} & \colhead{Group} & \colhead{Name} & $M_V$ & \colhead{Type} & \colhead{$T_\mathrm{eff}$ (K)} & \colhead{Air mass}
}
\decimalcolnumbers
\startdata
1 & A-f, B-f & HR2088 & 1.90 & A2IV & 8840 & 1.17\\
1 & A-l, B-l & HR2088 & 1.90 & A2IV & 8840 & 1.17\\
2 & A-f, B-f & HR4534 & 2.14 & A3V & 8550 & 1.11\\
2 & A-l, B-l & HR3982 & 1.35 & B7V & 14000 & 1.37\\
3 & A-f, B-f & HR0936 & 2.12 & B8V & 12500 & 1.26\\
3 & A-l, B-l & HR4534 & 2.14 & A3V & 8550 & 1.13\\
4 & A-f, B-f & HR4534 & 2.14 & A3V & 8550 & 1.09\\
4 & A-l, B-l & HR3982 & 1.35 & B7V & 14000 & 1.33\\
\enddata
\tablecomments{
Column (1): data number in this paper, corresponding to that in Table \ref{tab:obsinfo}. Column (2): data group where the spectrum of the standard star is used. Column (3): names of standard stars. Columns (4) and (5): $V$-band magnitude and spectral types of the standard stars, respectively \citep{Hoffleit1995}. Column (6): effective temperatures of the standard stars \citep{Pecaut2013}. Column (7): mean air mass of standard stars.
}
\end{deluxetable}

\section{Analysis}\label{sec:analysis}
This section explains the methods for placing continuum, removing emission lines, and excluding bad data points in Sections \ref{subsec:cont_place}, \ref{subsec:line_sub}, and \ref{subsec:bad_data}. Section \ref{subsec:analysis_model_fit} shares the methods for decomposing each CO rovibrational absorption line in the derived spectrum of IRAS08 NW.

\subsection{Continuum Placement}\label{subsec:cont_place}
\begin{figure}
    \centering
    \includegraphics[width=\linewidth]{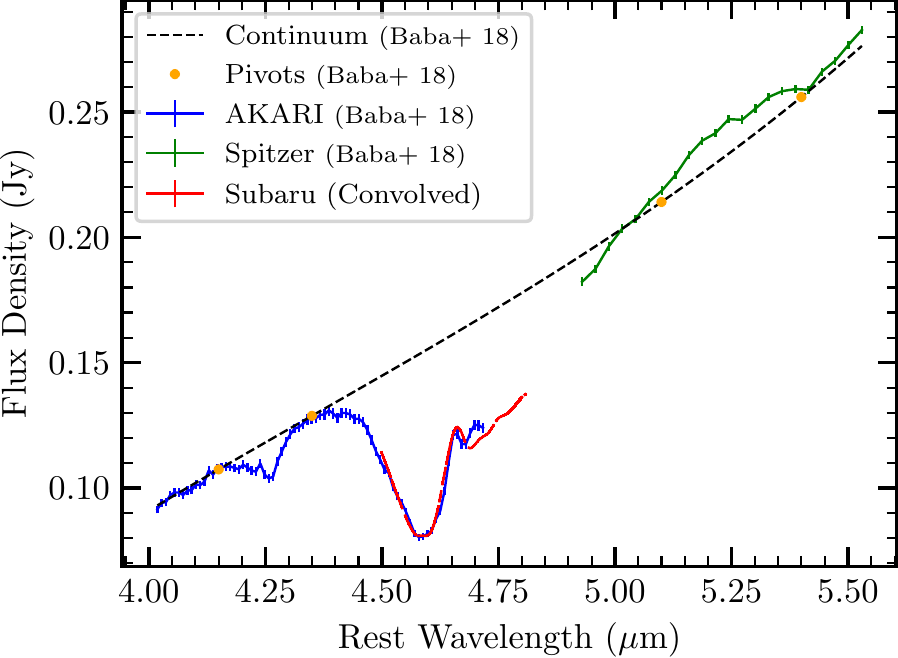}
    \caption{IRAS08 spectra derived from Subaru observation (this work) and AKARI/Spitzer observation \citep{Baba2018}. The flux of Subaru spectra is convolved and scaled to that of AKARI/Spitzer spectra.}
    \label{fig:scaletoAK}
\end{figure}

\begin{figure*}
    \centering
    \includegraphics[width=\linewidth]{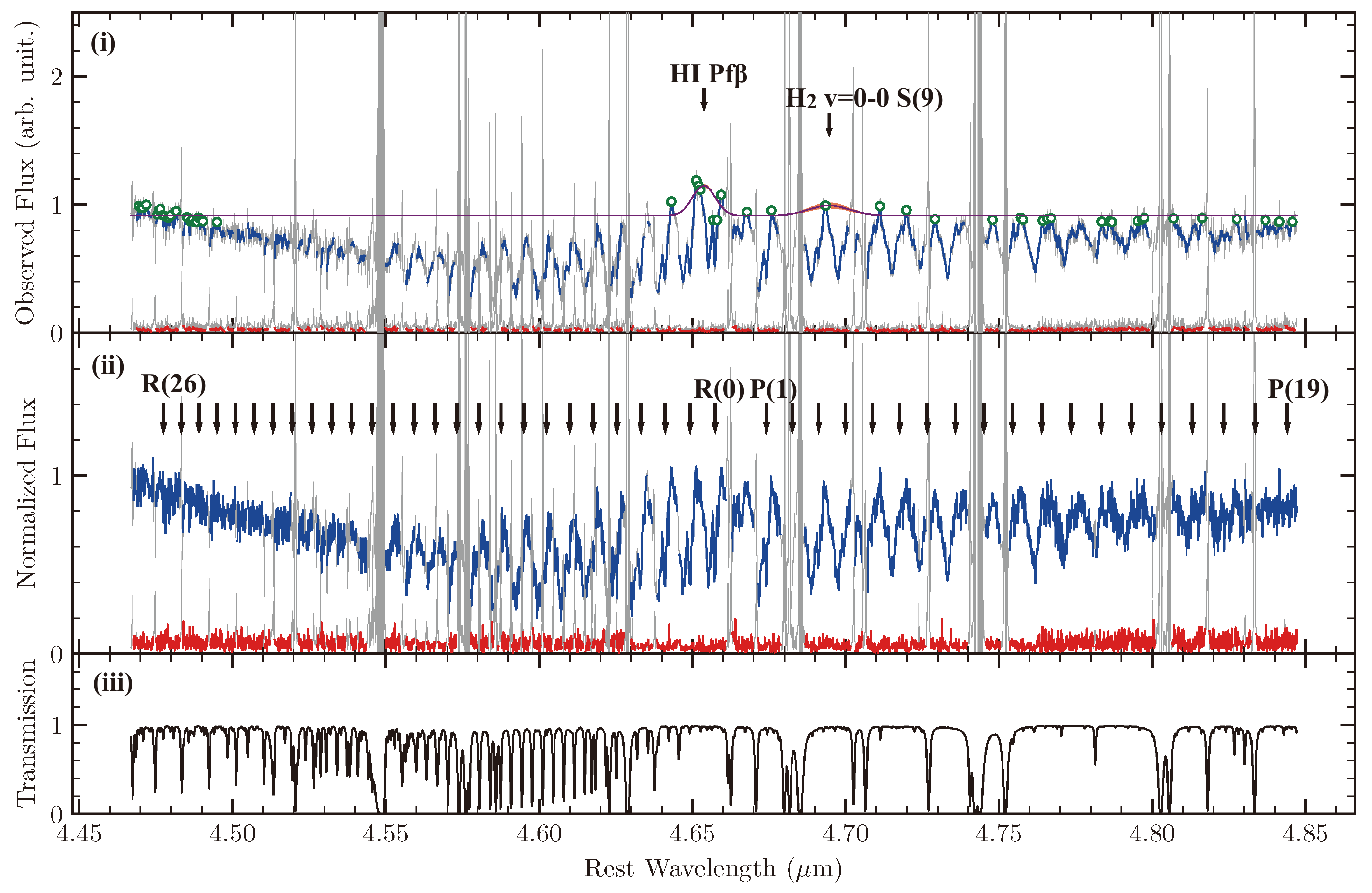}
    \caption{(i) Spectrum of IRAS08 NW normalized with the continuum determined by \citet{Baba2018}. The rest wavelength of the $\mathrm{H_I}$ Pf$\beta$ and $\mathrm{H_2}$ $v=0\to 0$ S(9) lines is denoted by the vertical arrows. The data points used to fit Gaussians to the two emission lines are denoted by the green-edged points. The best-fit model for the two emission lines is represented by a solid purple line. Its 1$\sigma$ confidence band is filled with orange. (The statistic error of the fitting is propagated.) (ii) Normalized spectrum of IRAS08 NW, with the emission lines subtracted. The continuum level is 1. The normalized flux of IRAS08 NW is shown as a solid blue line, while the flux error is shown as a solid red line. The excluded bad data points with the flux error of $\delta{F}<0.025$ or $0.2<\delta{F}$ or the telluric transmission of $T<0.7$ are colored in gray. The rest wavelength of the CO rovibrational absorption lines is denoted by the vertical arrows. (iii) Telluric transmission at the corresponding wavelength range corrected by the redshift of $z=0.0583$.}
    \label{fig:reducted_spec}
\end{figure*}

It is difficult to safely determine the continuum of the CO rovibrational absorption band because of the crowded lines distributed across the wavelength range observed with Subaru IRCS. Thus, we adopted the continuum determined by \citet{Baba2018} using low-resolution spectra with a wider wavelength range derived with AKARI ($R\sim 165$) and Spitzer ($R\sim 80$). To apply the AKARI/Spitzer continuum to our Subaru spectra, we scaled the flux levels of the Subaru spectra after convolving them and matching the wavelength resolution to that of the AKARI/Spitzer spectra. Figure \ref{fig:scaletoAK} shows the scaled Subaru spectra (this work; red points) and the AKARI/Spitzer spectra (\citealp{Baba2018}; blue/green points).
We then divided the scaled Subaru spectra by the continuum (\citealp{Baba2018}; black dashed line). Figure \ref{fig:reducted_spec}(i) shows the spectra divided by the continuum estimated from the AKARI spectra.

\subsection{Subtraction of Emission Lines}\label{subsec:line_sub}
We had to exclude the contribution from the two emission lines in this wavelength range: $\mathrm{H_I}$ Pf$\beta$ at $\lambda=4.65378\,\mathrm{\mu{m}}$ and $\mathrm{H_2}$ $v=0\to 0$ S(9) at $\lambda=4.69461\,\mathrm{\mu{m}}$. These lines were subtracted by fitting two Gaussian profiles to the spectrum peaks, with their central wavelength and width fixed. The central wavelength was fixed to the rest values. The line widths were fixed to those of $\mathrm{H_I}$ Br$\gamma$ and $\mathrm{H_2}$ $v=1\to 0$ S(1) of 241 and $430\,\mathrm{km\,s^{-1}}$ in the standard deviation, respectively, which were observed toward IRAS08 by \citet{Goldader1995}. In this paper, we denote these flux levels after subtracting the emission lines as the ``normalized flux.''

\subsection{Exclusion of Bad Data Points}\label{subsec:bad_data}
After the process in Sections \ref{subsec:cont_place} and \ref{subsec:line_sub}, we selected data points with $T>0.7$ and $0.025<\delta{F}<0.2$, where $T$ is the telluric transmission, and $\delta{F}$ is the error of the normalized flux, to exclude bad data points with too large or small flux errors. We then derived the final spectrum of the CO rovibrational absorption lines from which the contributions of the two emission lines and the bad data points were removed (Figure \ref{fig:reducted_spec}(ii)). The derived S/N was $\sim 20$ against the continuum level.

\begin{deluxetable}{llcccc}
\tablecaption{Wavelength, Oscillator Strength, and Lower-state energy of the ${}^{12}\mathrm{CO}\ (v=0\to 1,\ J\to J'=J\pm 1)$ Lines Considered in This Study
\label{tab:line_info}}
\tablehead{
\colhead{} & \colhead{} & \multicolumn2c{$R$-branch} & \multicolumn2c{$P$-branch}\\
\colhead{} & \colhead{} & \multicolumn2c{($J'=J+1$)} & \multicolumn2c{($J'=J-1$)}\\
\colhead{$J$} & \colhead{$E_J$} & \colhead{$\lambda_{JJ'}$} & \colhead{$f_{JJ'}$} & \colhead{$\lambda_{JJ'}$} & \colhead{$f_{JJ'}$}\\
\colhead{} & \colhead{(K)} & \colhead{$\mathrm{(\mu m)}$} & \colhead{($10^{-6}$)} & \colhead{$\mathrm{(\mu m)}$} & \colhead{($10^{-6}$)}
}
\startdata
0	&	0.0000	&	4.6575	&	11.6587	&	---	&	---	\\
1	&	5.5321	&	4.6493	&	7.7884	&	4.6742	&	3.8715	\\
2	&	16.5963	&	4.6412	&	7.0260	&	4.6826	&	4.6371	\\
3	&	33.1919	&	4.6333	&	6.7034	&	4.6912	&	4.9585	\\
4	&	55.3183	&	4.6254	&	6.5272	&	4.6999	&	5.1308	\\
5	&	82.9749	&	4.6177	&	6.4224	&	4.7088	&	5.2382	\\
6	&	116.1603	&	4.6100	&	6.3526	&	4.7177	&	5.3079	\\
7	&	154.8734	&	4.6024	&	6.3056	&	4.7267	&	5.3558	\\
8	&	199.1130	&	4.5950	&	6.2690	&	4.7359	&	5.3908	\\
9	&	248.8768	&	4.5876	&	6.2459	&	4.7451	&	5.4154	\\
10	&	304.1634	&	4.5804	&	6.2283	&	4.7545	&	5.4303	\\
11	&	364.9706	&	4.5732	&	6.2164	&	4.7640	&	5.4428	\\
12	&	431.2960	&	4.5662	&	6.2049	&	4.7736	&	5.4530	\\
13	&	503.1368	&	4.5592	&	6.1989	&	4.7833	&	5.4596	\\
14	&	580.4908	&	4.5524	&	6.1973	&	4.7931	&	5.4610	\\
15	&	663.3546	&	4.5456	&	6.1961	&	4.8031	&	5.4646	\\
16	&	751.7253	&	4.5389	&	6.1946	&	4.8131	&	5.4616	\\
17	&	845.5995	&	4.5324	&	6.1956	&	4.8233	&	5.4622	\\
18	&	944.9734	&	4.5259	&	6.1987	&	4.8336	&	5.4604	\\
19	&	1049.8433	&	4.5195	&	6.2004	&	4.8440	&	5.4567	\\
20	&	1160.2053	&	4.5132	&	6.2037	&	---	&	---	\\
21	&	1276.0551	&	4.5071	&	6.2083	&	---	&	---	\\
22	&	1397.3883	&	4.5010	&	6.2142	&	---	&	---	\\
23	&	1524.2002	&	4.4950	&	6.2212	&	---	&	---	\\
24	&	1656.4859	&	4.4891	&	6.2260	&	---	&	---	\\
25	&	1794.2405	&	4.4832	&	6.2316	&	---	&	---	\\
26	&	1937.4587	&	4.4775	&	6.2381	&	---	&	---	\\
\enddata
\tablecomments{
The oscillator strength was calculated from the Einstein $A$-coefficient. See the text for the details. $E_J,\ \lambda,$ and Einstein $A$-coefficients were derived from the high-resolution transmission molecular absorption database \citep{Coxon2004,Li2015,Gordon2017}.
}
\end{deluxetable}

\subsection{Velocity Decomposition of the CO Gas Lines}\label{subsec:analysis_model_fit}
\begin{figure}
    \centering
    \includegraphics[width=\linewidth]{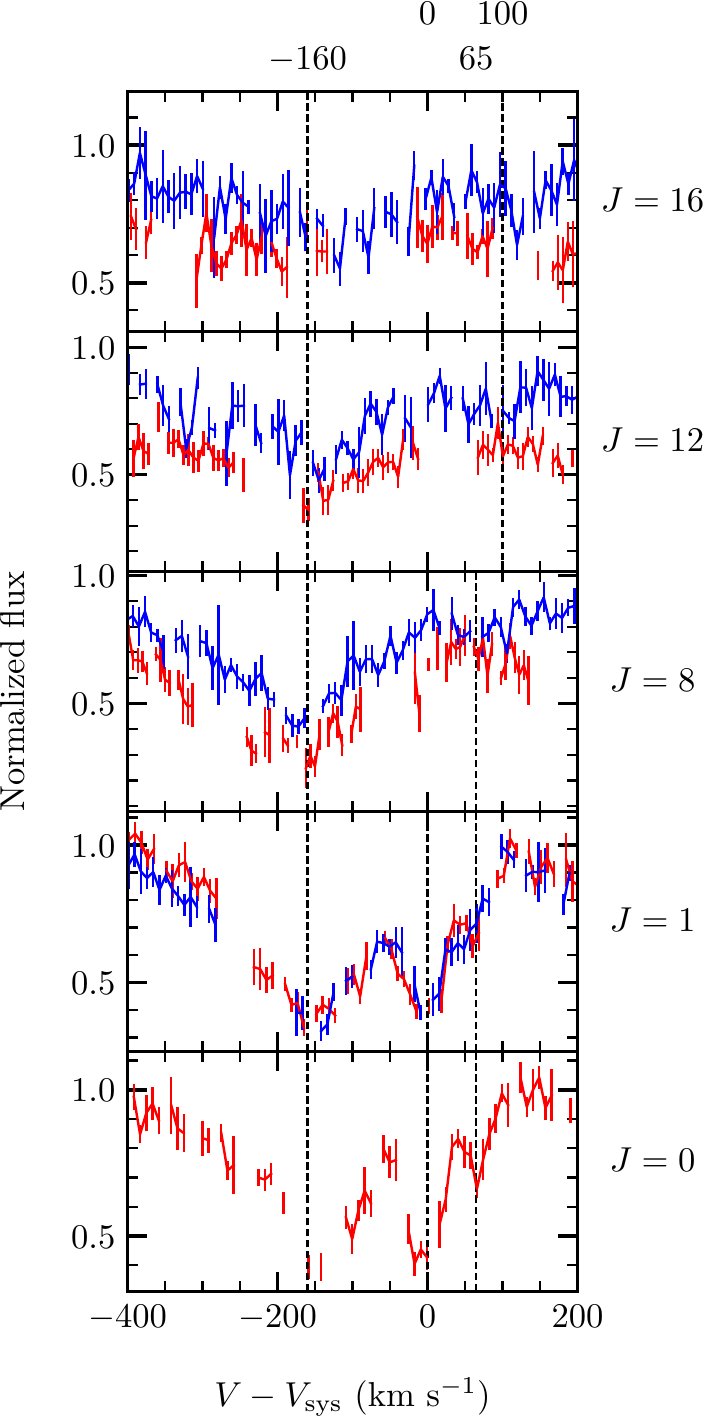}
    \caption{Spectrum of the $R$-branch (red) and $P$-branch (blue) at $J=0,\ 1,\ 8,\ 12,\ \text{and}\ 16$. The abscissa is the LOS velocity relative to the system. The ordinate is the normalized flux. The vertical dashed lines denote the position of $V-V_\mathrm{sys}=-160,\ 0,\ +65,\ \text{and}\ +100\,\mathrm{km\,s^{-1}}$.}
    \label{fig:vel_comp}
\end{figure}

Figure \ref{fig:vel_comp} shows the velocity profiles of some gaseous CO transitions in the spectrum derived in Sections \ref{subsec:cont_place}--\ref{subsec:bad_data}. The CO velocity profiles in Figure \ref{fig:vel_comp} showed some velocity components in the CO rovibrational absorption lines of IRAS08 NW, as suggested in the studies of \citet{Geballe2006a} and \citet{Shirahata2013b}.
In particular, the component with $V-V_\mathrm{sys}\sim -160\,\mathrm{km\,s^{-1}}$ appeared in all lines, while the other components were detected in some of the lines. The components of $V-V_\mathrm{sys}\sim 0$ and $\sim +65\,\mathrm{km\,s^{-1}}$ appeared in low rotational levels of $J\lesssim 8$ and are likely to be attributed to the absorbers of the low excitation temperature. On the contrary, the component of $V-V_\mathrm{sys}\sim +100\,\mathrm{km\,s^{-1}}$ appeared in high rotational levels of $J\gtrsim 12$ and is likely to be attributed to the absorber of the high excitation temperature.\footnote{We did not attribute the redshifted components to more blueshifted components of the adjacent CO transitions because the line widths were narrower than the component at $V-V_\mathrm{sys}\sim -160\,\mathrm{km\,s^{-1}}$, indicating that they were located farther from the central black hole and should have smaller absolute values of the LOS velocity.}\par
We then fitted some Gaussian profiles to the optical depth of the CO absorption lines to decompose their velocity components using Lmfit v1.0.0 packages \citep{Lmfit}. The absorption transitions from $v=0$ to $v=1$ were assumed to be dominant, and the emission transitions were assumed to be negligible. These assumptions are justified because the temperature of the gas detected as absorption ($T_\mathrm{gas}$) should be less than that of the NIR source, which is the dust sublimation layer of the temperature of $T_\mathrm{sub}\sim 1500\,\mathrm{K}$ \citep{Netzer1993}, while the energy difference between the CO vibrational levels of $v=0$ and $v=1$ is typically $T_{v01}\sim 2000\,\mathrm{K}$, or $T_\mathrm{gas}(<T_\mathrm{sub})<T_{v01}$. Then, the optical depth $\tau(\lambda)$ can be expressed as $\tau(\lambda)=-\ln(F_\lambda/F_\mathrm{c})$ with the normalized flux $F_\lambda/F_\mathrm{c}$, where $F_\mathrm{c}$ is the continuum flux. In this work, we assumed that the NIR light source was fully covered with absorbers. In other words, an area covering factor was assumed to be unity. If this assumption is not the case and the covering fraction is smaller, the estimated CO column density becomes larger by a factor of $\lesssim 7$. Moreover, the estimated kinetic temperature becomes higher by $\lesssim 25\%$ at the most extreme case, as discussed in Appendix \ref{app:area_cov}. These differences in the parameter estimation do not affect the conclusion of this paper.\par
The optical depth $\tau(\lambda)$ of each absorption line, $R(J)$ ($v=0\to 1,\ J\to J+1$) and $P(J)$ ($v=0\to 1,\ J\to J-1$), can be written as the sum of the optical depth $\tau_i(\lambda)$ of the $i$th component in the absorption lines as
\begin{linenomath}
\begin{gather}
    \tau(\lambda)=\sum_i\tau_{i}(\lambda).\label{eq:sum_tau}
\end{gather}
\end{linenomath}
The optical depth $\tau_i(\lambda)$ of each component is expressed with the column density $N_{J,i}$ of the CO molecules at $v=0,\ J$ as
\begin{linenomath}
\begin{gather}
    \tau_i(\lambda)=\frac{\pi e^2}{m_\mathrm{e} c^2}N_{J,i}f_{JJ'}\lambda_{JJ'}^2\phi_i(\lambda),\label{eq:opt_depth}\\
    \phi_i(\lambda)=\frac{1}{\sqrt{2\pi}\sigma_{\lambda,i}}\exp\left[-\frac{(\lambda-\lambda_{0,i})^2}{2\sigma_{\lambda,i}^2}\right],\\
    \lambda_{0,i}=\left(1+\frac{V_{0,i}}{c}\right)\lambda_{JJ'},\ \sigma_{\lambda,i}=\frac{\sigma_{V,i}}{c}\lambda_{JJ'},\label{eq:vel2lam}
\end{gather}
\end{linenomath}
where $\pi e^2/m_\mathrm{e}c^2=8.8523\times 10^{-13}\,\mathrm{cm}$; $f_{JJ'}$ is the oscillator strength of the transition; $\lambda_{0,i}$ and $\sigma_{\lambda,i}$ are the central wavelength and the standard deviation of the $i$th component in the absorption line, respectively; $V_{0,i}$ and $\sigma_{V,i}$ are the velocity centroid and the velocity width of the $i$th component, respectively; and $\lambda_{JJ'}$ is the rest wavelength of the transition. We assumed herein a Gaussian profile as a line profile function $\phi(\lambda)$ because the Einstein $A$-coefficients of these transitions were at most $\sim 35\,\mathrm{s^{-1}}$, and the FWHMs of the natural broadening were less than $\sim 2\times 10^{-12}\,\mathrm{\mu{m}}$, which was negligible compared to those of the observed absorption lines. The oscillator strength of $v=0\to 1,\ J\to J'$, $f_{JJ'}$, wa calculated from the Einstein $A$-coefficients of $v=1\to 0,\ J'\to J$, $A_{J'J}$, as follows \citep{Goorvitch1994}:
\begin{linenomath}
\begin{gather}
    f_{JJ'}=\frac{2J'+1}{2J+1}\frac{m_\mathrm{e}c}{8\pi^2 e^2}\lambda_{JJ'}^2A_{J'J}.
\end{gather}
\end{linenomath}
Table \ref{tab:line_info} shows the rest wavelength and the oscillator strength of the absorption lines used in this work.
In the fitting, we determined $\lambda_{0,i}$ and $\sigma_{\lambda,i}$ from $V_{0,i}$ and $\sigma_{V,i}$, respectively. In short, the free parameters were $V_{0,i}$ and $\sigma_{V,i}$ for the $i$th component, and $N_{J,i}$ for each rotational level $J$ of the $i$th component.\par
In addition, we also introduced ice features to reproduce $P(1)$ and $R(1)$ lines, whose optical depth at the peaks was almost equal. The oscillator strength of $R(1)$ was approximately twice as great as that of $P(1)$; thus, the optical depth of $R(1)$ should be twice as great as that of $P(1)$ according to Equation (\ref{eq:opt_depth}). However, they had an almost equal optical depth, and there have to be other absorption features under $P(1)$. In this wavelength range around the band center of $4.66\text{--}4.68\,\mathrm{\mu{m}}$, we can find two apolar CO ice absorption bands \citep{Boogert2015b}. Appendix \ref{sec:ice_feature} explains the details of the ice features.\par
\begin{figure}
    \centering
    \includegraphics[width=0.95\linewidth]{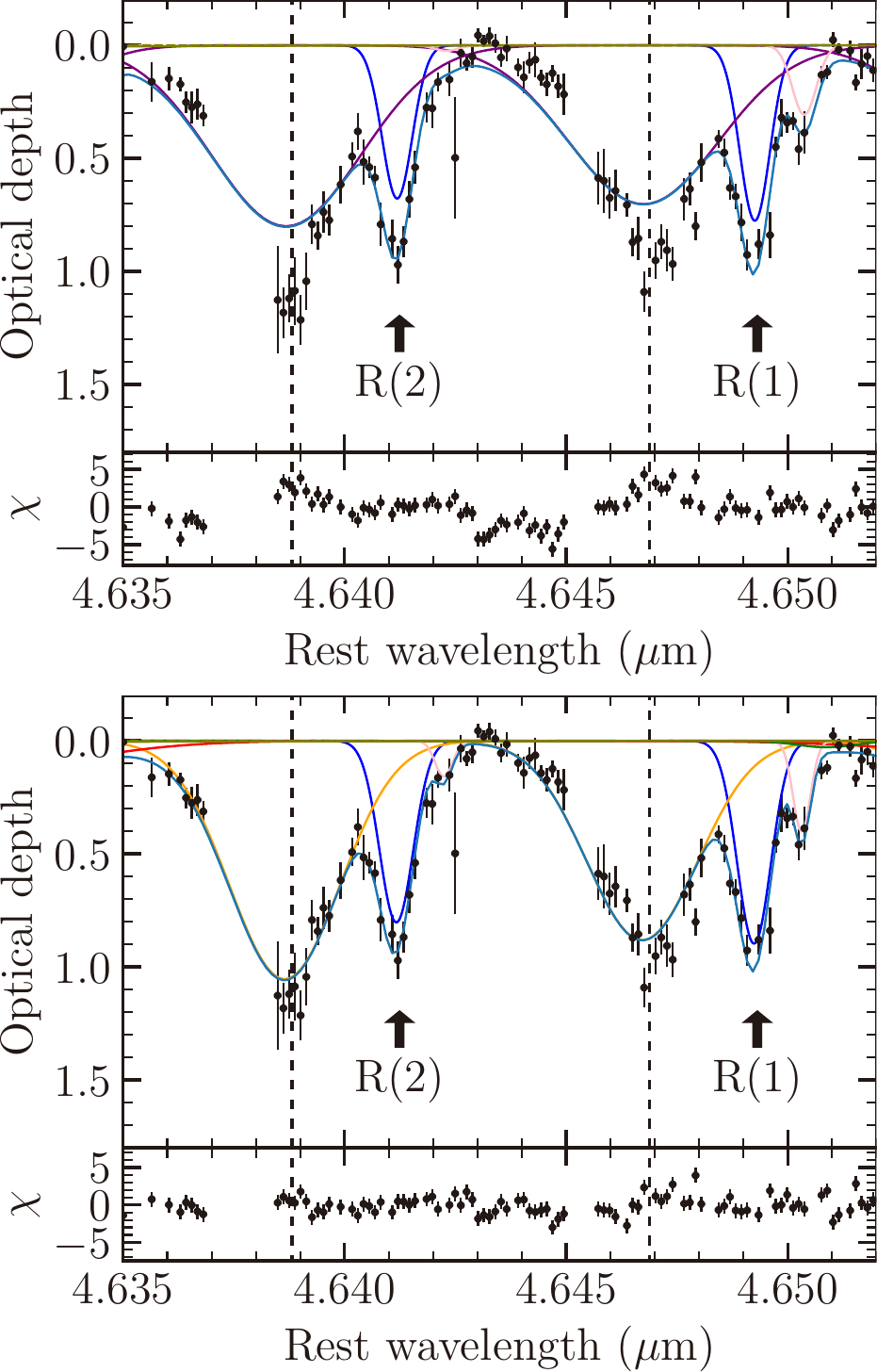}
    \caption{Top: best-fit model when the velocity component around $V-V_\mathrm{sys}\sim -160\,\mathrm{km\,s^{-1}}$ is assumed to consist of one Gaussian component at the upper half and the residual significance at the lower half. The components with the velocity centroids of $V-V_\mathrm{sys}\sim +65,\ 0,\ \text{and}\ -160\,\mathrm{km\,s^{-1}}$ are shown as the solid lines colored in pink, blue, and purple, respectively. Their sum is shown as a sky-blue solid line. For the excess around the absorption peaks to be clear, the ordinates are the optical depth. The wavelengths corresponding to the velocities of $V-V_\mathrm{sys}=-160\,\mathrm{km\,s^{-1}}$ in $R(1)$ and $R(2)$ lines are denoted by the vertical dashed lines. Bottom: best-fit model when the velocity component around $V-V_\mathrm{sys}\sim -160\,\mathrm{km\,s^{-1}}$ is assumed to consist of two Gaussian components at the upper half and the residual significance at the lower half. The narrower and broader components around $V-V_\mathrm{sys}\sim -160\,\mathrm{km\,s^{-1}}$ are shown as solid lines colored in orange and red. We should note that the narrower component is dominant in $J=1$ and 2 and the broader component does not appear in the transitions in this figure. The components with the velocity centroids of $V-V_\mathrm{sys}\sim +65\ \text{and}\ 0\,\mathrm{km\,s^{-1}}$ are shown as solid lines colored in pink and blue, respectively. Their sum is shown as a sky-blue solid line.}
    \label{fig:excess_lowj}
\end{figure}

As demonstrated in Figure \ref{fig:excess_lowj}, even if we fitted Gaussian profiles to each component, excess was detected around the peaks at low rotational levels.
This indicates that low-$J$ absorption lines have narrower components of $V-V_\mathrm{sys}\sim -160\,\mathrm{km\,s^{-1}}$ than high-$J$ absorption lines. Therefore, we assumed that this component consisted of narrow and broad components, and the narrow one dominated in the lower $J$, whereas the broad one dominated in higher $J$. We fitted this component with two Gaussian profiles having different widths. This process improved the $\chi^2$ value from $\chi_\nu^2=4305/2067$ to $\chi_\nu^2=3705/2052$, with the difference being $\Delta\chi^2=600$ having the degree-of-freedom difference of $\Delta\nu=15$. The $\Delta\chi^2$ value rejected the model describing the velocity component of $V-V_\mathrm{sys}\sim -160\,\mathrm{km\,s^{-1}}$ as one component (Figure \ref{fig:excess_lowj}, top panel), with less than 0.1\% significance level. In summary, the additional component that could not be visually inspected was detected around $V-V_\mathrm{sys}\sim -160\,\mathrm{km\,s^{-1}}$.\par
\begin{figure*}
    \centering
    \includegraphics[width=0.95\linewidth]{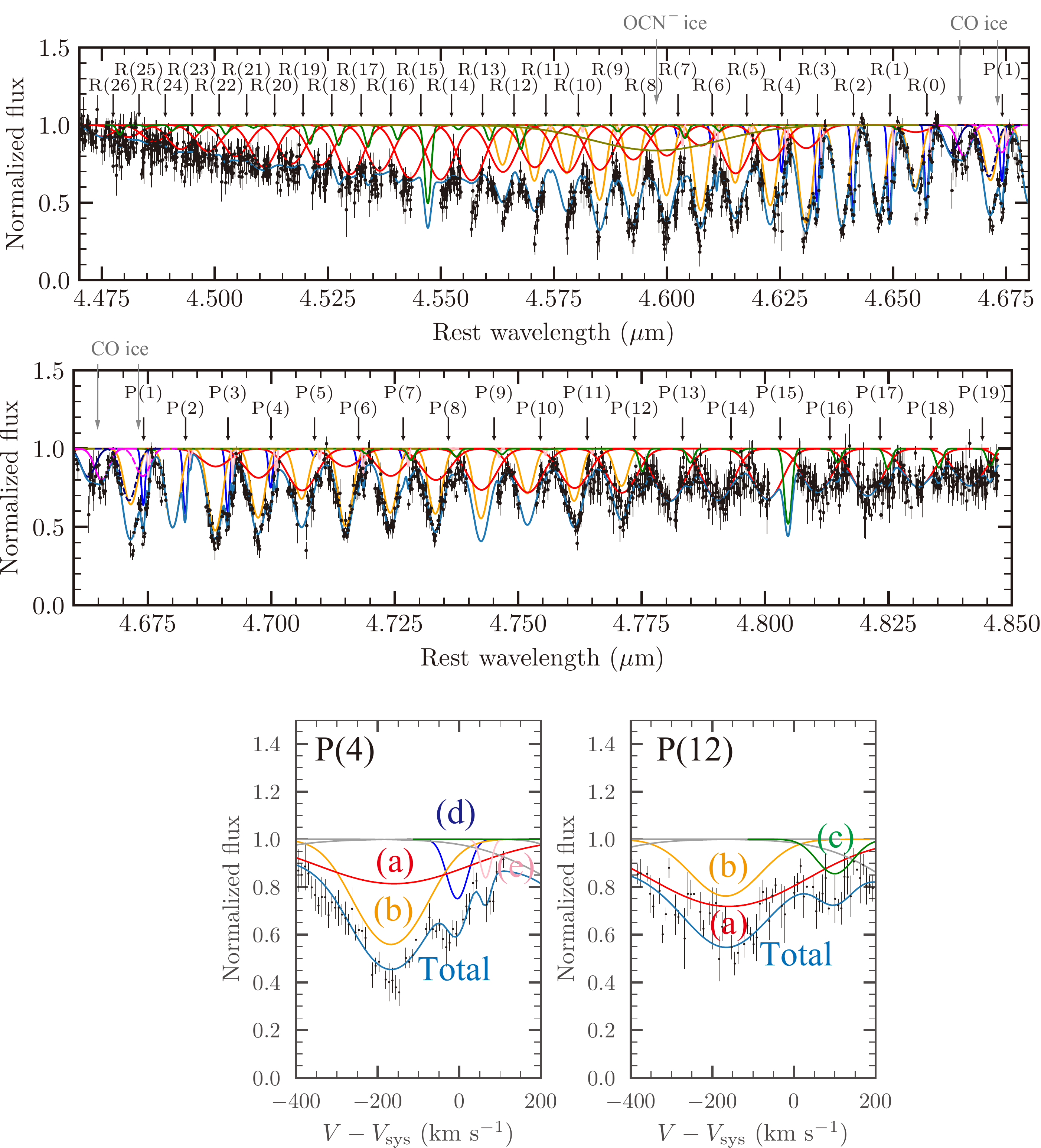}
    \caption{Top: best-fit model of the gaseous CO rovibrational absorption lines in IRAS08 NW. The abscissa is the rest wavelength. The ordinate is the normalized flux. Components (a)--(e) and their sum are denoted in red, orange, green, dark blue, pink, and sky-blue colors, respectively. The black arrows denote the rest wavelength of each CO transition. The absorption of the apolar $\mathrm{OCN^-}$ ice (4.598 $\mathrm{\mu{m}}$) is represented by the olive-green solid line. In addition, the pure apolar $\mathrm{CO}$ ice (4.665 $\mathrm{\mu{m}}$) and the $\mathrm{CO_2}$-mixed apolar $\mathrm{CO}$ ice (4.673 $\mathrm{\mu{m}}$) are shown with solid and dashed lines, respectively. Their red- and blueshifted components are colored in magenta and navy blue, respectively. See Appendix \ref{sec:ice_feature} for the details of the detected ice features. Bottom: components (a)--(e) and their sum detected in the gaseous ${}^{12}\mathrm{CO}$ $P(4)$ and $P(12)$ absorption lines. The abscissa is the LOS velocity relative to the systemic velocity ($V-V_\mathrm{sys}$). The ordinate is the normalized flux. Each component and the sum are colored as in the left panels. The components of the other lines than $P(4)$ and $P(12)$ are colored in gray.}\label{fig:fitres_colines}
\end{figure*}

Finally, we found five components, i.e., $i=\text{(a)--(e)}$ in Equations (\ref{eq:sum_tau})--(\ref{eq:vel2lam}), in each CO rovibrational absorption line, which are illustrated in Figure \ref{fig:fitres_colines}.
Table \ref{tab:fitres_dynst_cdens} presents the estimated velocity centroid, velocity deviation, and column density $N_J$ of components (a)--(e). As for component (c), the velocity centroid and the velocity dispersion were fixed to the values determined from the visual inspections of the peak positions and the FWHMs because they could not be determined with the fitting.\par
Moreover, each component was detected at different ranges of rotational levels. Component (a), which had a blueshifted velocity centroid of $V-V_\mathrm{sys}\sim -160\,\mathrm{km\,s^{-1}}$ and the largest velocity dispersion of $\sigma_V\sim 175\,\mathrm{km\,s^{-1}}$ of all components, was detected in all rotational levels of $0\le J\le 26$, indicating a high excitation temperature. Component (b), which had a similar velocity centroid and a small velocity dispersion compared to component (a), was detected in relatively low rotational levels of $0\le J\le 12$, indicating a lower excitation temperature than that of component (a). Component (c), which had a redshifted velocity centroid of $V-V_\mathrm{sys}\sim +100\,\mathrm{km\,s^{-1}}$ and a smaller velocity dispersion than components (a) and (b), was detected in relatively high rotational levels of $6\le J\le 26$, indicating a higher excitation temperature than that of component (a). Component (d), which had a systemic velocity centroid and a smaller velocity dispersion than components (a)--(c), was detected in low rotational levels of $0\le J\le 7$, indicating a low excitation temperature. Component (e), which had a redshifted velocity centroid of $V-V_\mathrm{sys}\sim +65\,\mathrm{km\,s^{-1}}$ and the smallest velocity dispersion of all components, was also detected in low rotational levels of $0\le J\le 9$, indicating a low excitation temperature.\par
For each component, we also set 3$\sigma$ lower limits to ${}^{12}\mathrm{CO}/{}^{13}\mathrm{CO}$ abundance ratios by fitting each velocity component of ${}^{13}\mathrm{CO}$ rovibrational absorption lines ($v=0\to 1,\ \Delta{J}=\pm{1}$) with the inverse ratio (${}^{13}\mathrm{CO}/{}^{12}\mathrm{CO}$) free and the velocity centroid ($V_0$) and dispersion ($\sigma_V$) identical to those of ${}^{12}\mathrm{CO}$ simultaneously. The rest wavelength and the Einstein $A$-coefficients of ${}^{13}\mathrm{CO}$ transitions, which were necessary to calculate the oscillator strength, were derived from the high-resolution transmission molecular absorption database \citep{Coxon2004,Li2015,Gordon2017} as in ${}^{12}\mathrm{CO}$. Table \ref{tab:fitres_dynst_cdens} summarizes the results. In all components, the ratios did not reject selective dissociation \citep[e.g.,][]{vanDishoeck1988}.

\begin{deluxetable*}{llccccc}
\tablecaption{Estimated Velocity Centroid ($V_0$), Velocity Dispersion ($\sigma_V$), CO Column Densities at $v=0,\ J$ ($N_J$), and ${}^{12}\mathrm{CO}/{}^{13}\mathrm{CO}$ Ratio of Components (a)--(e)
\label{tab:fitres_dynst_cdens}}
\tablehead{
\colhead{} & \colhead{} & \colhead{(a)} & \colhead{(b)} & \colhead{(c)} & \colhead{(d)} & \colhead{(e)}
}
\startdata
$V_0\,(\mathrm{km\,s^{-1}})$ & {} & $-160\pm 5$ & $-167\pm 2$ & 100 (fix) & $-5\pm 1$ & $65\pm 1$\\
$\sigma_V\,(\mathrm{km\,s^{-1}})$ & {} & $175\pm 4$ & $80\pm 3$ & 42 (fix) & $24\pm 1$ & $13\pm 1$\\
$N_J\,(10^{17}\,\mathrm{cm^{-2}})$ & $J=0$ & $\le$ 0.90 	&	 0.71 $\pm$ 0.18 	&	 $\le$ 0.09	&	 0.30 $\pm$ 0.04 	&	 0.07 $\pm$ 0.01 	\\
{} & $J=1$ & $\le$ 0.90 	&	 1.85 $\pm$ 0.20 	&	 $\le$ 0.12	&	 0.57 $\pm$ 0.04 	&	 0.13 $\pm$ 0.02 	\\
{} & $J=2$ & $\le$ 1.09 	&	 2.46 $\pm$ 0.26 	&	 $\le$ 0.15	&	 0.57 $\pm$ 0.05 	&	 0.06 $\pm$ 0.02 	\\
{} & $J=3$ & 0.86 $\pm$ 0.51 	&	 2.42 $\pm$ 0.32 	&	 $\le$ 0.17	&	 0.51 $\pm$ 0.05 	&	 0.12 $\pm$ 0.02 	\\
{} & $J=4$	&	 1.41 $\pm$ 0.33 	&	 1.83 $\pm$ 0.23 	&	 $\le$ 0.17	&	 0.28 $\pm$ 0.05 	&	 0.09 $\pm$ 0.02 	\\
{} & $J=5$	&	 2.08 $\pm$ 0.32 	&	 1.19 $\pm$ 0.24 	&	 $\le$ 0.14	&	 0.05 $\pm$ 0.04 	&	 0.03 $\pm$ 0.02 	\\
{} & $J=6$	&	 0.80 $\pm$ 0.32 	&	 2.07 $\pm$ 0.22 	&	 0.05 $\pm$ 0.05	&	 0.12 $\pm$ 0.03 	&	 0.09 $\pm$ 0.03 	\\
{} & $J=7$	&	 1.25 $\pm$ 0.33 	&	 1.58 $\pm$ 0.20 	&	 0.12 $\pm$ 0.06	&	 0.06 $\pm$ 0.06 	&	 0.06 $\pm$ 0.03 	\\
{} & $J=8$	&	 1.45 $\pm$ 0.28 	&	 1.60 $\pm$ 0.19 	&	 0.08 $\pm$ 0.04	&	 $\le$ 0.09 	&	 $\le$ 0.06 	\\
{} & $J=9$	&	 1.97 $\pm$ 0.30 	&	 1.75 $\pm$ 0.33 	&	 0.06 $\pm$ 0.04	&	 \ldots 	&	 0.03 $\pm$ 0.02 	\\
{} & $J=10$	&	 2.11 $\pm$ 0.25 	&	 0.99 $\pm$ 0.22 	&	 $\le$ 0.14	&	 \ldots 	&	 $\le$ 0.07 	\\
{} & $J=11$	&	 1.85 $\pm$ 0.21 	&	 1.19 $\pm$ 0.16 	&	 $\le$ 0.19	&	 \ldots 	&	 $\le$ 0.12 	\\
{} & $J=12$	&	 2.10 $\pm$ 0.20 	&	 0.79 $\pm$ 0.15 	&	 0.24 $\pm$ 0.04	&	 \ldots 	&	 \ldots 	\\
{} & $J=13$	&	 2.58 $\pm$ 0.10 	&	 \ldots &	 0.14 $\pm$ 0.04	&	 \ldots 	&	 \ldots 	\\
{} & $J=14$	&	 2.54 $\pm$ 0.09 	&	 \ldots 	&	 0.04 $\pm$ 0.04	&	 \ldots 	&	 \ldots 	\\
{} & $J=15$	&	 2.45 $\pm$ 0.11 	&	 \ldots 	&	 $\le$ 16.92	&	 \ldots 	&	 \ldots 	\\
{} & $J=16$	&	 2.06 $\pm$ 0.10 	&	 \ldots 	&	 0.14 $\pm$ 0.04	&	 \ldots 	&	 \ldots 	\\
{} & $J=17$	&	 2.25 $\pm$ 0.09 	&	 \ldots 	&	 0.22 $\pm$ 0.05	&	 \ldots 	&	 \ldots 	\\
{} & $J=18$	&	 1.75 $\pm$ 0.09 	&	 \ldots 	&	 0.20 $\pm$ 0.04	&	 \ldots 	&	 \ldots 	\\
{} & $J=19$	&	 1.69 $\pm$ 0.08 	&	 \ldots 	&	 0.18 $\pm$ 0.05	&	 \ldots 	&	 \ldots 	\\
{} & $J=20$	&	 1.86 $\pm$ 0.15 	&	 \ldots 	&	 $\le$ 0.17	&	 \ldots 	&	 \ldots 	\\
{} & $J=21$	&	 1.33 $\pm$ 0.11 	&	 \ldots 	&	 $\le$ 0.15	&	 \ldots 	&	 \ldots 	\\
{} & $J=22$	&	 1.10 $\pm$ 0.12 	&	 \ldots 	&	 0.09 $\pm$ 0.05	&	 \ldots 	&	 \ldots 	\\
{} & $J=23$	&	 1.00 $\pm$ 0.11 	&	 \ldots 	&	 0.07 $\pm$ 0.05	&	 \ldots 	&	 \ldots 	\\
{} & $J=24$	&	 0.67 $\pm$ 0.10 	&	 \ldots 	&	 0.05 $\pm$ 0.04	&	 \ldots 	&	 \ldots 	\\
{} & $J=25$	&	 0.47 $\pm$ 0.09 	&	 \ldots 	&	 $\le$ 0.11	&	 \ldots 	&	 \ldots 	\\
{} & $J=26$	&	 0.45 $\pm$ 0.08 	&	 \ldots 	&	 0.09 $\pm$ 0.05	&	 \ldots 	&	 \ldots 	\\
${}^{12}\mathrm{CO}/{}^{13}\mathrm{CO}$ & {} & $\ge 7.5$ & $\ge 19.6$ & $\ge 2.9$ & $\ge 13.2$ & $\ge 4.3$
\enddata
\tablecomments{
The upper limits are 3$\sigma$ upper limits. Ellipsis dots indicate the column densities with which the fittings did not converge.
}
\end{deluxetable*}

\section{The Origin of Each Component}\label{sec:origin_comp}
Section \ref{subsec:analysis_model_fit} showed that the five discrete components with different LOS velocities or velocity dispersions were detected in each CO rovibrational absorption line. These components are then likely to have originated from some different structure. This section discusses the location, excitation mechanism, and physical properties of each component.

\subsection{Location}\label{subsec:dist_radii}
We have two potential ways to estimate the location of each component: (1) the excitation temperature assuming the central heating, and (2) the velocity dispersion assuming the dynamics driven by the central black hole.\par
We first check the validity of option 1. The excitation temperature based on the level population of $v=0$ is subjected to the FIR-to-(sub)millimeter radiation fields (\citealp{Maloney1994}; Matsumoto et al. in preparation), which are expected to be ubiquitous in the central parsec-scale regions of the AGN. Hence, the excitation temperature can be different from the gas kinetic temperature.\par
We then check the validity of option 2. If we adopt the black hole mass of $M_\mathrm{BH}=9\times 10^7M_\odot$ based on the $H$-band luminosity \citep{Veilleux2002,Veilleux2009} and the stellar velocity dispersion of the host galaxy of $\sigma_*=180\,\mathrm{km\,s^{-1}}$ based on the $M_\mathrm{BH}$-$\sigma_*$ relation \citep{Tremaine2002}, the radius of the sphere of influence in IRAS08 NW is $GM_\mathrm{BH}/\sigma_*^2\sim 12\,\mathrm{pc}$. Thus, the dynamics in the molecular torus, whose size is expected to be a few parsecs, is supposed to be driven by the central black hole. We then estimate the location of each component based on the velocity dispersion assuming the dynamics driven by the central black hole.\par
In this work, we assume that the clump dynamics is the sum of the Kepler rotation, turbulence, and inflowing or outflowing motion. If the molecular torus is assumed to be a hydrostatic disk, as in many previous theoretical studies \citep[e.g.,][]{Beckert2004,Vollmer2004,Hopkins2012}, the ratio of the rotating velocity ($V_\mathrm{rot}$) to the velocity dispersion ($\sigma_V$) is similar to that of the rotating radius ($R_\mathrm{rot}$) to the disk height ($H$), or $\sigma_V/V_\mathrm{rot}\sim H/R_\mathrm{rot}$. In addition, assuming that the molecular torus is the triangle disk with a constant $H/R_\mathrm{rot}$ ratio, or $H/R_\mathrm{rot}\sim \text{const.}$, we can assume that the ratio of the velocity dispersion to the rotating velocity is constant, or (i) $\sigma_V/V_\mathrm{rot}\sim \text{const.}$ Because the dynamics is supposed to be driven by the central black hole, the rotating velocity is related to the radius as (ii) $V_\mathrm{rot}\propto R_\mathrm{rot}^{-0.5}$. Then, assumptions (i) and (ii) lead the relationship between the rotating radius and the velocity dispersion of $R_\mathrm{rot}\propto\sigma_V^{-2}$, and we can determine the rotating radius of each component based on it. Although the assumption of the torus disk with the constant $\sigma_V/V_\mathrm{rot}$ ratio is based on the hydrostatic disk as the first step, it is also applicable to a hydrodynamic radiation fountain model \citep{Wada2016}, which predicts that the $V_\mathrm{rot}/\sigma_V$ ratio is nearly constant in the inner region of the torus. In addition, the theoretical models, such as the CLUMPY \citep{Nenkova2008a} and XCLUMPY \citep{Tanimoto2019} models, which assume the triangle disk, well reproduce SEDs of AGNs in Seyfert galaxies.\par
For the above reasons, the ratio of the rotating radii of component (b) to component (a) is
\begin{linenomath}
    \begin{gather}
        \frac{R_\mathrm{rot,b}}{R_\mathrm{rot,a}}=\left(\frac{\sigma_{V,\mathrm{b}}}{\sigma_{V,\mathrm{a}}}\right)^{-2}=4.8\pm 0.4.\label{eq:ratio_radii_compab}
    \end{gather}
\end{linenomath}
In the same manner, the ratios of the rotating radii of the other components to component (a) are
\begin{linenomath}
    \begin{align}
        \frac{R_\mathrm{rot,c}}{R_\mathrm{rot,a}}&\approx 17,\\
        \frac{R_\mathrm{rot,d}}{R_\mathrm{rot,a}}&= 53\pm 5,\label{eq:ratio_compd}\\
        \intertext{and}
        \frac{R_\mathrm{rot,e}}{R_\mathrm{rot,a}}&= 180\pm 30\label{eq:ratio_compe}.
    \end{align}
\end{linenomath}
In the MHD torus model of \citet{Chan2017}, the power-law index for the rotation velocity was predicted to be in the range of $-1.0$ to $-0.84$, which is steeper than that of the Kepler rotation of $-0.5$, in the most stable configuration. Thus, the above ratios based on the assumption of the Kepler rotation may be the upper limits.
Here, component (d) is attributed to the host galaxy because the LOS velocity is the systemic velocity, and the velocity dispersion is smaller than those in components (a)--(c), giving the large ratio of the rotating radius to component (a). In addition, the low excitation temperature of component (d) mentioned in Section \ref{subsec:analysis_model_fit} is consistent with this. Component (e) is also attributed to the host galaxy because the velocity dispersion is smaller than the host galactic component (d). Although the LOS velocity of component (e) indicates an infalling motion, we do not discuss its origin because the feature is too narrow to be resolved with the spectral resolution of $R\sim 10{,}000$. Here, components (d) and (e) are out of the sphere of influence; thus, their ratios in Equations (\ref{eq:ratio_compd}) and (\ref{eq:ratio_compe}) can be different from the true values. Thus, we focus on components (a)--(c), attributing them to the molecular torus in this paper. Refer to Appendix \ref{app:host_gal} for the details of components (d) and (e).\par
\begin{figure*}
    \centering
    \includegraphics[width=0.75\linewidth]{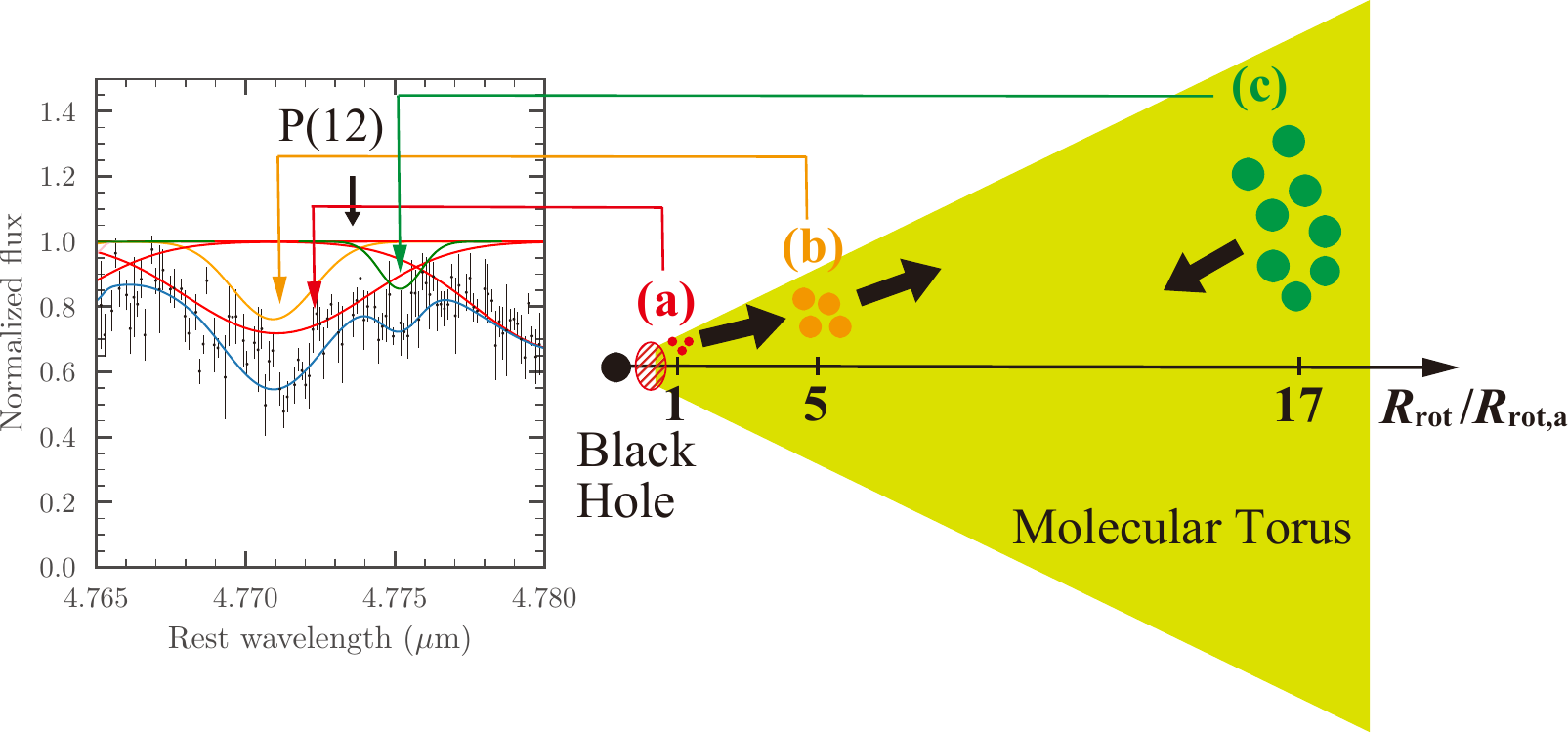}
    \caption{Sketch of the geometry of components (a)--(c) in the molecular torus. Components (d) and (e) are excluded here because they are unlikely to be related to the molecular torus according to their small velocity dispersions of $\sigma_V\sim 13\text{--}24\,\mathrm{km\,s^{-1}}$}
    \label{fig:torus_geom}
\end{figure*}

The velocity dispersions of each component indicate that the rotating radii of components (a)--(c) obey
\begin{linenomath}
\begin{gather}
    R_\mathrm{rot,a}<R_\mathrm{rot,b}<R_\mathrm{rot,c}.
\end{gather}
\end{linenomath}
Therefore, we assume the geometry of each component in the torus as shown in Figure \ref{fig:torus_geom}.
To be observed as absorption lines, these components have to be located in front of the dust sublimation layer, which is expected to be the NIR source as mentioned in Section \ref{sec:intro}. Hence, the innermost component (a) should be located farther from the black hole than the dust sublimation layer. If we assume an isotropic central radiation, the dust sublimation radius is expected to be $R^\mathrm{iso}_\mathrm{sub}\approx 1\,\mathrm{pc}$ in IRAS08 NW ($L_\mathrm{AGN}\approx 9\times 10^{45}\,\mathrm{erg\,s^{-1}}$; \citealp{Efstathiou2014}) according to \citet{Barvainis1987}. More realistically, the radius is the upper limit because the ultraviolet (UV) radiation from the accretion disk is anisotropic and weaker toward the equatorial direction than toward the polar direction.\par
We evaluate the average radius of the dust sublimation layer as follows to roughly estimate the rotating radius of component (a):
\begin{linenomath}
    \begin{align}
        \begin{aligned}
            \overline{R^\mathrm{ani}_\mathrm{sub}}&\approx\frac{\int^{\pi/2}_{\Theta_\mathrm{h}} R^\mathrm{iso}_\mathrm{sub}\sqrt{\cos\theta(1+2\cos\theta)/3}\sin\theta\mathrm{d}\theta}{\int^{\pi/2}_{\Theta_\mathrm{h}} \sin\theta\mathrm{d}\theta}\\
            &\approx 0.5\,\mathrm{pc},
        \end{aligned}
    \end{align}
\end{linenomath}
where $\theta$ and $\Theta_\mathrm{h}=2\pi/9\,(=40^\circ)$ are the polar angle and the polar half-opening angle of the torus estimated by the model fitting to SED of IRAS08 NW \citep{Vega2008}, respectively, and $\sqrt{\cos\theta(1+2\cos\theta)/3}$ is the anisotropy of the dust sublimation radius predicted by \citet{Netzer1987}.
If we assume that the innermost component (a) is located near the dust sublimation layer and the rotating radius is $R_\mathrm{rot,a}\approx 0.5\,\mathrm{pc}$, the rotating radii of components (b) and (c) are expected to be $R_\mathrm{rot,b}\approx 2\,\mathrm{pc}$ and $R_\mathrm{rot,c}\approx 8\,\mathrm{pc}$. In short, the velocity centroids and dispersions of each component agree with the dynamical structure of the torus, where the molecular clouds are outflowing in the inner regions and inflowing in the outer regions (Figure \ref{fig:torus_geom}).

\subsection{Excitation Mechanism and Physical Properties}\label{subsec:each_comp}
After determining the column density of the CO molecules at each rotational level $J$, we can now estimate the temperature and the column density of each absorber attributed to each component in the absorption lines based on the level population. Based on the detections of warm ($200\text{--}500\,\mathrm{K}$) CO gas with a large column density ($N_\mathrm{H}\gtrsim 10^{23}\,\mathrm{cm^{-2}}$) in nearby AGNs, the molecular torus is likely to be heated by the X-ray radiation from the central region of the AGN \citep{Baba2018} because the high-energy X-ray photons can penetrate into the torus medium more deeply than UV photons according to X-ray-dominated region (XDR) models \citep{Maloney1996,Meijerink2005}. In this section, we investigate the excitation mechanism and the physical properties of components (a)--(c) assuming the central X-ray heating.

\subsubsection{Component (a)}\label{subsec:comp_a}
\begin{figure*}
    \includegraphics[width=0.9\linewidth]{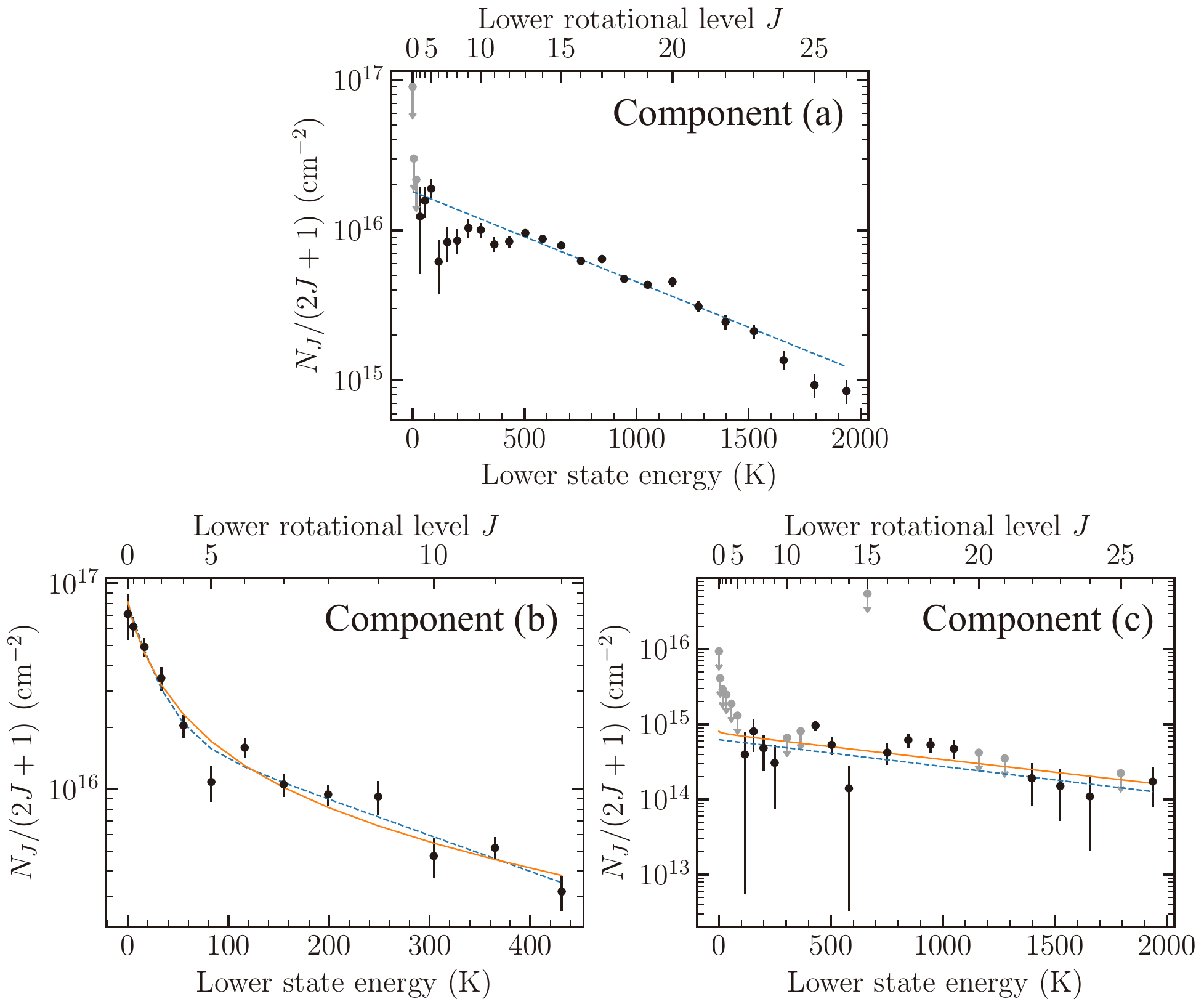}
    \caption{Population diagrams of components (a)--(c). The blue dashed lines are the best-fit Boltzmann distributions. In components (b) and (c), the orange solid lines are the non-LTE level population calculated with RADEX. Some parameters are not well constrained in components (b) and (c); thus, the median values of the posterior distributions of each parameter shown in Figures \ref{fig:mcmchist_comp2} and \ref{fig:mcmchist_comp5} are used to draw the orange solid lines. The column densities, whose upper limits are set, are colored in gray.}
    \label{fig:boltz_plot_torus}
\end{figure*}

As illustrated in Section \ref{subsec:dist_radii}, component (a) is the innermost component of components (a)--(c). Figure \ref{fig:boltz_plot_torus} depicts the population diagram of this component.
Its level population is in the LTE up to the rotational level of $J= 23$, where the lower-state energy is $E_J=1524\,\mathrm{K}$ because $\log[N_J/(2J+1)]$ in $J\le 23$ is well aligned on a line, which is the Boltzmann distribution in Figure \ref{fig:boltz_plot_torus}. Under the LTE condition, the level population of the CO molecules can be described by the Boltzmann distribution and written as
\begin{linenomath}
\begin{gather}
    \frac{N_{J}}{2J+1}=\frac{N_{\mathrm{CO}}}{Z(T_{\mathrm{ex}})}\exp\left(-\frac{E_J}{k_\mathrm{B}T_{\mathrm{ex}}}\right),
\end{gather}
\end{linenomath}
where $N_{\mathrm{CO}}$ is the total column density of the CO molecules in all the rotational levels of $v=0$ in the component; $T_{\mathrm{ex}}$ is the excitation temperature; $E_J$ is the rotational energy of the energy level of $v=0,\ J$, and $Z(T_{\mathrm{ex}})$ is the partition function calculated with the Total Internal Partition Sums (TIPS; \citealp{Gamache2017}) code. Thus, we fit the Boltzmann distribution to the observed $N_{J}$ of each $J$ level to determine $T_{\mathrm{ex}}$ and $N_{\mathrm{CO}}$ of component (a) using the Lmfit v1.0.0 package \citep{Lmfit}. Table \ref{tab:line_info} shows the rotational energy $E_J$ of each $J$ level used in this work.\par
The estimated excitation temperature and the CO molecular column density are $T_\mathrm{ex,a}=721\pm 43\,\mathrm{K}$ and $N_\mathrm{CO,a}=(4.8\pm 0.1)\times 10^{18}\,\mathrm{cm^{-2}}$, respectively. Figure \ref{fig:boltz_plot_torus} shows the best-fit Boltzmann distribution as a dashed line. Table \ref{tab:res_mcmc_compbe} summarizes the estimated physical properties.\par
In addition, the hydrogen molecular density should be larger than the critical density of the rotational level of $J=23$ for the level population in $J\le 23$ to be in LTE. Here, the critical density of $J$ is defined as
\begin{linenomath}
    \begin{gather}
        n_\mathrm{cr}(J)=\frac{\sum_{J^\prime<J}A_{v=0,JJ^\prime}}{\sum_{J^\prime\ne J}k_{v=0,JJ^\prime}(T_\mathrm{kin})}\label{eq:crit_dens}
    \end{gather}
\end{linenomath}
where $A_{v=0,JJ^\prime}$ and $k_{v=0,JJ^\prime}$ are the Einstein $A$-coefficient and the collisional rate coefficient for $v=0,\ J\to J^\prime$ transition, respectively \citep[e.g.,][]{Osterbrock2006}. Under the LTE condition, the excitation temperature is equal to the kinetic temperature ($T_\mathrm{ex,a}=T_\mathrm{kin,a}$). The critical density of $J=23$ is $n_\mathrm{cr}(J=23)\approx 2\times 10^6\,\mathrm{cm^{-3}}$ for $T_\mathrm{kin,a}=721\pm 43\,\mathrm{K}$. Here, the collisional rate coefficients are referred from \citet{Yang2010} via the Leiden Atomic and Molecular Database (LAMDA; \citealp{Schoier2005a}). Note that only the CO transitions between the energy levels in $v=0,\ J\le 40$ are considered, and the ortho-to-para ratio of the hydrogen molecules is assumed to be thermal \citep{Burton1992}. Thus, the hydrogen molecular density of component (a) is $n_\mathrm{H_2,a}\gtrsim 2\times 10^6\,\mathrm{cm^{-3}}$. An XDR model predicts that the fractional abundance of the CO molecules is $\sim 10^{-4}$ at the temperature of $T\lesssim 700\,\mathrm{K}$ in the dense gas of $n_\mathrm{H}\sim 10^5\,\mathrm{cm^{-3}}$ \citep{Maloney1996}; therefore, the dense and hot clumps of component (a) reasonably exist.\par
Based on the CO column density and the lower limit of the hydrogen molecular density, we can impose the upper limit on the geometrical thickness of component (a) along the LOS considering a volume filling factor of the clumps ($\phi_V$) as
\begin{linenomath}
    \begin{align}
    \begin{aligned}
        d_\mathrm{los,a}&\sim \frac{N_\mathrm{H_2,a}}{n_\mathrm{H_2,a}\phi_V}= \frac{[\mathrm{H_2}]}{[\mathrm{CO}]}\frac{N_\mathrm{CO,a}}{n_\mathrm{H_2,a}\phi_V}\\
        &\lesssim 0.3\left(\frac{\phi_V}{0.03}\right)^{-1}\,\mathrm{pc},
    \end{aligned}
    \end{align}
\end{linenomath}
where we assume the abundance ratio of the CO molecules to the $\mathrm{H_2}$ molecules as $[\mathrm{CO}]/[\mathrm{H_2}]\sim 10^{-4}$ \citep{Dickman1978}. Here, the volume filling factor is set to $\phi_V\sim 0.03$ as a typical value according to some theoretical models of the molecular torus \citep[e.g.,][]{Beckert2004,Vollmer2004,Honig2007}. The estimated thickness is consistent with the torus size, which is expected to be a few parsecs. Thus, we attribute component (a) to the hot and dense clumps of $T_\mathrm{kin,a}=721\pm 43\,\mathrm{K}$ and $\log(n_\mathrm{H_2,a}/\mathrm{cm^{-3}})\gtrsim 6.3$ at the innermost region of the molecular torus of components (a)--(c).

\subsubsection{Component (b)}\label{subsec:comp_b}
As illustrated in Section \ref{subsec:dist_radii}, component (b) is located between components (a) and (c). Figure \ref{fig:boltz_plot_torus} presents the population diagram. We then find that its level population is folded at $E_J\sim 100\,\mathrm{K}$; thus, we consider two scenarios as the origin of the folded level population of component (b).

\paragraph{Temperature Gradient in an LTE Clump.}
The first candidate of the origin of component (b) is the temperature gradient in a clump. Some studies \citep[e.g.,][]{Nenkova2008a, Namekata2014} predicted that clumps in the AGN torus are illuminated by radiation from the central region and that a temperature gradient exists inside the clump, where the illuminated surface is the hottest and the shaded surface is the coldest. We then assume that the folded level population is the sum of two Boltzmann distributions with different excitation temperatures and estimate the physical properties of the cold and hot parts in analogy to the method in Section \ref{subsec:comp_a}. Consequently, the excitation temperature and the CO column density of the cold part are $T^\mathrm{cold}_\mathrm{ex,b}=22\pm 6\,\mathrm{K}$ and $N^\mathrm{cold}_\mathrm{CO,b}=(5\pm 1)\times 10^{17}\,\mathrm{cm^{-2}}$, respectively, while those of the hot part are $T^\mathrm{hot}_\mathrm{ex,b}=248\pm 41\,\mathrm{K}$ and $N^\mathrm{hot}_\mathrm{CO,b}=(1.8\pm 0.1)\times 10^{18}\,\mathrm{cm^{-2}}$, respectively.\par
We then check whether these parameters are reasonable by comparing them with the clumpy torus model of \citet{Nenkova2008a}. They modeled the dust temperature gradient inside a clump based on the equilibrium between the blackbody-emitted energy from it and the absorbed energy from the central radiation. Moreover, they found that the dust temperature dropped very sharply near the illuminated surface and the temperature was almost constant at the rest of the clump with an optical thickness of $\tau_V\gtrsim 10$. Because IRAS08 NW is likely to be observed in the nearly edge-on position \citep{Vega2008} and the silicate feature at $\lambda\sim 9.7\,\mathrm{\mu{m}}$ is observed as an absorption, clumps in the molecular torus should have a $V$-band optical thickness of $\tau_V\gtrsim 20$ \citep{Nenkova2008b}. In addition, the large optical depth is also supported from the observed column density. The CO column density of component (b) is $N_\mathrm{CO,b}=N^\mathrm{cold}_\mathrm{CO,b}+N^\mathrm{hot}_\mathrm{CO,b}\sim 2\times 10^{18}\,\mathrm{cm^{-2}}$ and equivalent to the column density of the hydrogen atoms of $N_\mathrm{H,b}\sim 4\times 10^{22}\,\mathrm{cm^{-2}}$ if we assume the abundance ratio of the CO molecules to the hydrogen molecules as $[\mathrm{CO}]/[\mathrm{H_2}]\sim 10^{-4}$ \citep{Dickman1978} and $N_\mathrm{H}=2N_\mathrm{H_2}$. Thus, the equivalent $V$-band optical depth is $\tau_V\sim 20\,(>10)$, which is determined using $A_V\approx N_\mathrm{H}/(1.8\times 10^{21}\,\mathrm{cm^{-2}})$ \citep{Bohlin1978a} and $\tau_V=A_V/1.086$, where $A_V$ is the $V$-band extinction.\par
Hence, if the two different temperatures are attributed to the temperature gradient in a clump of such a large $\tau_V$, the colder part of a clump should be optically thicker than the hotter part. However, the observed column density of the colder part is three times smaller than that of the hotter part. For the above reasons, this scenario for the origin of component (b) is not suitable for reproducing its energy distribution.

\paragraph{Non-LTE Radiative Excitation by Hot Background Radiation.}
The second candidate is clumps illuminated by the hot background radiation field from the far-infrared (FIR) to (sub)millimeter wavelength, which excites CO molecules to high $J$ levels in $v=0$. The NIR ($\lambda\sim \text{1--5}\,\mathrm{\mu{m}}$) source in the AGN is expected to be the thermal radiation from the hot dust of the temperature of $\gtrsim 1200\,\mathrm{K}$ by some studies (e.g., \citealp{Rees1969a,Rieke1978,Landt2011a}; Matsumoto et al. in preparation). Thus, the clumps near the dust sublimation layer in the molecular torus may be illuminated by the hot dust radiation. In illuminated clumps, the level population in high rotational levels, where the density in a clump is lower than the critical densities, is expected to be determined by radiative excitation, while that in low rotational levels is expected to be determined by collisional excitation.\par
Therefore, we check whether there are reasonable physical parameters, such as the kinetic temperature ($T_\mathrm{kin}$), the volume density of hydrogen molecules ($n_\mathrm{H_2}$), the brightness temperature of FIR-to-(sub)millimeter background radiation field ($T_\mathrm{bg}$), and the column density of CO ($N_\mathrm{CO}$), to reproduce the level population of component (b) with a statistical equilibrium radiative transfer code, RADEX v08sep2017 \citep{vanderTak2007}, assuming non-LTE clumps. Here, note that RADEX only considers the transitions between the energy levels with $v=0,\ J\le 40$. The RADEX model of the CO level population is fitted to the observed population of component (b) with the Markov Chain Monte Carlo (MCMC) method using Emcee v3.0.2 \citep{Foreman-Mackey2013} and Lmfit packages to investigate reasonable solutions whose distributions in parameter spaces are far from the normal distributions.\par
The free parameters are $T_\mathrm{kin}$, $n_\mathrm{H_2}$, $T_\mathrm{bg}$, and $N_\mathrm{CO}$. The prior distribution of each parameter is a box function. We assume herein the following conditions:
\begin{enumerate}[label=(\arabic*)]
    \item The kinetic temperature and the background brightness temperature ($T_\mathrm{kin}$, $T_\mathrm{bg}$) are lower than the dust sublimation temperature ($1500\,\mathrm{K}$) and higher than the cosmic microwave background (CMB) temperature ($2.73\,\mathrm{K}$). The prior distributions of the kinetic temperature and the background brightness temperature are the box functions in the range of $T_\mathrm{kin}$ and $T_\mathrm{bg}\in [2.73, 1500]$.
    \item The hydrogen molecular density ($n_\mathrm{H_2}$) is smaller than a typical value in a maser disk ($10^{10}\,\mathrm{cm^{-3}}$), which is expected to be located in the dense inner region of the torus \citep{Taniguchi1998}. Then, the prior distribution of the hydrogen molecular density is the box function in the range of $\log\qty(n_\mathrm{H_2}/\mathrm{cm^{-3}})\in [3, 10]$.
    \item The prior distribution of the CO column density is the box function in the range of $\log\qty(N_\mathrm{CO}/\mathrm{cm^{-2}})\in [15, 19]$.
\end{enumerate}
\par
The posterior distribution $p(\vb*{w}|\vb*{X})$ is
\begin{linenomath}
\begin{gather}
    p(\vb*{w}|\vb*{X})\propto p(\vb*{X}|\vb*{w})p(\vb*{w}),\\
    p(\vb*{X}|\vb*{w})\propto \exp\qty[-\frac{1}{2}\sum_{J}\qty(\frac{X^J-f(J;\vb*{w})}{\delta{X^J}})^2]
\end{gather}
\end{linenomath}
where $\vb*{w}$ and $p(\vb*{w})$ are a vector of the four free parameters mentioned above and their prior distribution, respectively; $\vb*{X}$ and $\vb*{\delta X}$ are the vectors of the observed column densities of the excited CO molecules in each rotational level divided by their statistical weight ($N_J/{2J+1}$) and their errors ($\delta N_J/{2J+1}$), respectively; and $f(J;\vb*{w})$ is the RADEX model function of $N_J/{2J+1}$ given the physical parameters of $\vb*{w}$. Here, 100 chains, each of which is 50,000 long, are generated with the burn-in length of 1000, such that we can generate a chain whose length is at least 50 times longer than the integrated autocorrelating steps of each free parameter ($<300$).\par
\begin{figure}
    \centering
    \gridline{\fig{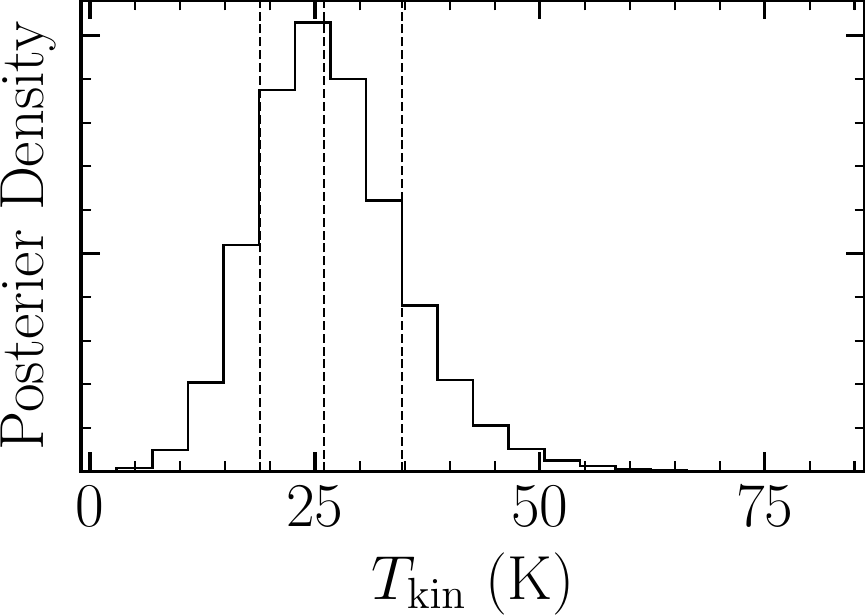}{0.45\linewidth}{}
              \fig{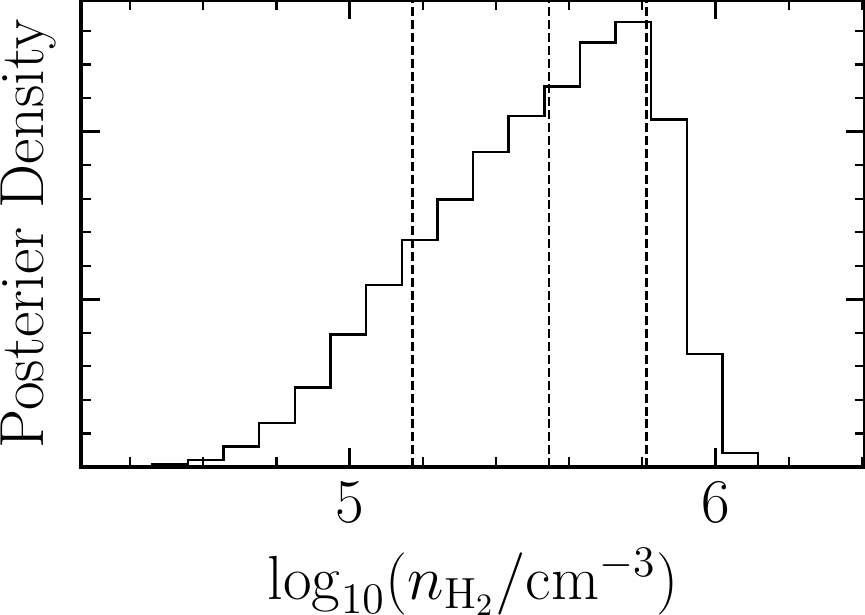}{0.45\linewidth}{}}
    \gridline{\fig{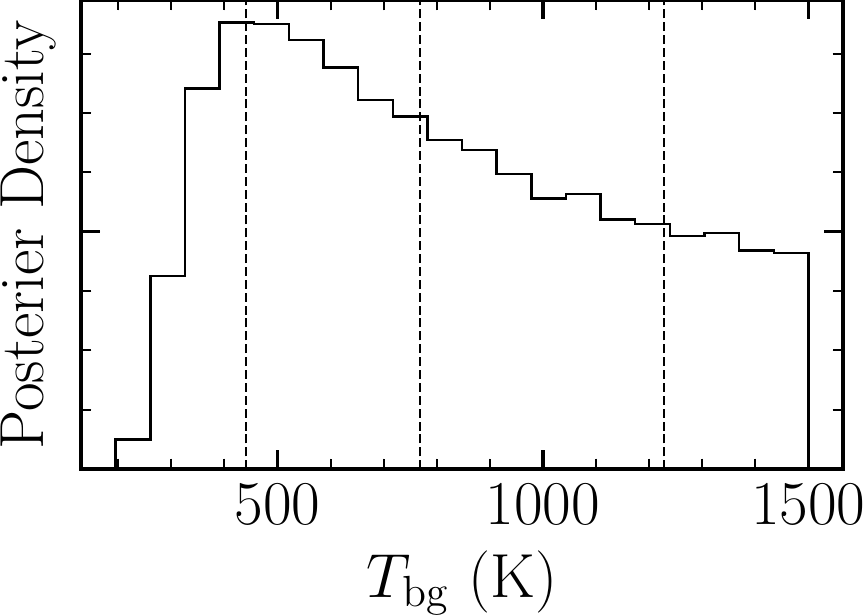}{0.45\linewidth}{}
              \fig{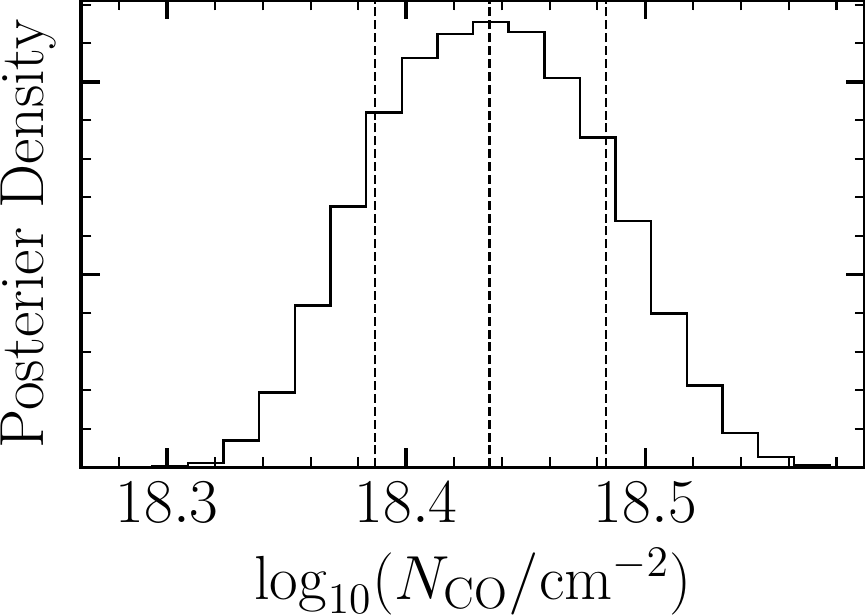}{0.45\linewidth}{}}
    \caption{Posterior distributions of $T_\mathrm{kin}$, $\log\qty(n_\mathrm{H_2}/\mathrm{cm^{-3}})$, $T_\mathrm{bg}$, and $\log\qty(N_\mathrm{CO}/\mathrm{cm^{-2}})$ of component (b). The dashed vertical lines indicate 16th, 50th, and 84th percentiles. The prior boundaries of each parameter are $[2.73, 1500],\ [3, 10],\ [2.73, 1500],\ \text{and}\ [15, 19]$. $T_\mathrm{bg}$ is cut off at $1500\,\mathrm{K}$ because of the boundary; hence, only the lower limit can be imposed.}\label{fig:mcmchist_comp2}
\end{figure}

Figure \ref{fig:mcmchist_comp2} shows the posterior distributions of estimated parameters. Table \ref{tab:res_mcmc_compbe} summarizes the estimated parameters.
Figure \ref{fig:boltz_plot_torus} depicts the non-LTE level population of component (b) with the median values of the posterior distributions. The non-LTE scenario can reproduce the observed level population; thus, we reject the LTE scenario and attribute component (b) to the clumps with a kinetic temperature of $T_\mathrm{kin,b}=26^{+9}_{-7}\,\mathrm{K}$, density of $\log\qty(n_\mathrm{H_2,b}/\mathrm{cm^{-3}})=5.5^{+0.3}_{-0.4}$, and CO column density of $\log\qty(N_\mathrm{CO,b}/\mathrm{cm^{-2}})=18.43\pm 0.05$. Only the lower limit is imposed on the brightness temperature of $T_\mathrm{bg,b}\ge 236\,\mathrm{K}$.\par
In analogy to component (a), the geometrical thickness of component (b) is estimated as
\begin{linenomath}
\begin{align}
\begin{aligned}
    d_\mathrm{los,b}&\sim \frac{N_\mathrm{H_2,b}}{n_\mathrm{H_2,b}}= \frac{[\mathrm{H_2}]}{[\mathrm{CO}]}\frac{N_\mathrm{CO,b}}{n_\mathrm{H_2,b}\phi_V}\\
    &\sim 0.7\left(\frac{\phi_V}{0.03}\right)^{-1}\,\mathrm{pc},
\end{aligned}
\end{align}
\end{linenomath}
where we assume the abundance ratio of CO to $\mathrm{H_2}$ molecules to be $[\mathrm{CO}]/[\mathrm{H_2}]\sim 10^{-4}$ \citep{Dickman1978} and the volume filling factor to be $\phi_V\sim 0.03$, as in Section \ref{subsec:comp_a}. The estimated thickness is consistent with the torus size, expected to be a few parsecs; hence, the non-LTE scenario does not violate the consistency of the geometrical size. We therefore suggest that component (b) is attributed to the absorbing clumps of the kinetic temperature of $T_\mathrm{kin}=26^{+9}_{-7}\,\mathrm{K}$ and the moderate density of $\log\qty(n_\mathrm{H_2}/\mathrm{cm^{-3}})=5.5^{+0.3}_{-0.4}$ illuminated by the strong radiation whose brightness temperature is $\gtrsim 236\,\mathrm{K}$.

\begin{deluxetable}{lccc}
\tablecaption{Estimated Physical Properties of Components (a)--(c), Which Are Attributed to the Molecular Torus\label{tab:res_mcmc_compbe}}
\tablehead{
\colhead{} & \colhead{(a)} & \colhead{(b)} & \colhead{(c)}
}
\startdata
LTE/NLTE & LTE & NLTE & NLTE\\
$T_\mathrm{kin}\,(\mathrm{K})$ & $721\pm 43$ & $26^{+9}_{-7}$ & \ldots\\
$\log (n_\mathrm{H_2}/\mathrm{cm^{-3}})$ & $\gtrsim 6.3$ & $5.5_{-0.4}^{+0.3}$ & $\le 5.5$\\
$T_\mathrm{bg}\,(\mathrm{K})$ & \ldots & $\ge 236$ & $\ge 784$\\
$\log (N_\mathrm{CO}/\mathrm{cm^{-2}})$ & $18.68\pm 0.02$ & $18.43\pm 0.05$ & $17.53\pm 0.04$
\enddata
\tablecomments{Component (a) is likely to be in the LTE, while the others are likely to be in the non-LTE (NLTE).}
\end{deluxetable}

\subsubsection{Component (c)}\label{subsec:comp_c}
As illustrated in Section \ref{subsec:dist_radii}, component (c) is located in the outermost region of components (a)--(c) and inflowing with the LOS velocity of $V-V_\mathrm{sys}\sim +100\,\mathrm{km\,s^{-1}}$. Figure \ref{fig:boltz_plot_torus} presents the population diagram. Although it is not clear whether or not the level population is in the LTE, the excitation temperature and the CO column density are estimated as $T_\mathrm{ex,c}=1218\pm 421\,\mathrm{K}$ and $N_\mathrm{CO,c}=(3.0\pm 0.8)\times 10^{17}\,\mathrm{cm^{-2}}$ based on the LTE assumption in analogy to Section \ref{subsec:comp_a}. This component at least has the highest excitation temperature because the component is observed only in high energy levels of $6\le J\le 26$ (Figure \ref{fig:vel_comp}). However, the situation where the gas temperature is the highest at the farthest point from the black hole is not reasonable because the molecular torus is predicted to be heated by the X-ray radiation from the central region of the AGN \citep{Baba2018} and the gas temperature of the outermost component (c) should be the lowest of components (a)--(c). To reproduce such a peculiar situation, we consider two scenarios for the origin of component (c).
\paragraph{Shock Heating Rather Than X-Ray Heating.}
Shock heating is a possible candidate of the heating process in the location distant from the central black hole. There are two types of shocks, namely, J-shock (with the discontinuity) and C-shock (without the discontinuity) \citep{Draine1980,Draine1981}. The temperature and the column density that can be reproduced are different \citep{McKee1984}. The shock types can be distinguished by the shock velocity and the pre-shock gas density because a fast shock is attained after the $\mathrm{H_2}$ dissociation within the shock wave, and it causes a rapid increase in the neutral gas temperature resulting J-shock \citep{Draine1983,Smith1990,LeBourlot2002}. Given the observed inflowing velocity along the LOS of $V_\mathrm{los,c}\approx 100\,\mathrm{km\,s^{-1}}$ and sufficient hydrogen molecular density for the levels up to $J=26$ to be in the LTE of $n_\mathrm{H_2,c}>n_\mathrm{cr}(J=26)\approx 2\times 10^6\,\mathrm{cm^{-3}}$ at the kinetic temperature of $T\sim 1200\,\mathrm{K}$ as defined in Equation (\ref{eq:crit_dens}), the possible shock in IRAS08 NW should be a J-shock \citep{LeBourlot2002}. Thus, we focus herein on J-shocks.\par
Some studies investigated the chemistry and the heating and cooling processes in J-shocks \citep[e.g.,][]{McKee1984,Hollenbach1989,Neufeld1989}. In the high-density regime with the pre-shock density of $n_\mathrm{H}\gtrsim 10^3\,\mathrm{cm^{-3}}$ and the shock velocity of $V_\mathrm{s}\sim 100\,\mathrm{km\,s^{-1}}$, the cooling processes between 5000 K and 100 K are dominated by the rotational transitions of OH, $\mathrm{H_2O}$, and CO and gas-grain collisions, and the column density of the hydrogen atoms from the shock front at $T=1000\,\mathrm{K}$ is given by $N_\mathrm{H}\approx 10^{20.5}\,\mathrm{cm^{-2}}$ \citep{McKee1984}. However, the 3$\sigma$ lower limit of the observed column density of the hydrogen atoms in component (c) is $N_\mathrm{H}\gtrsim 1.2\times 10^{21}\,\mathrm{cm^{-2}}\,(> 10^{20.5}\,\mathrm{cm^{-2}})$ if we assume $\mathrm{[CO]/[H_2]}\sim 10^{-4}$ and $N_\mathrm{H}=2N_\mathrm{H_2}$. Thus, we conclude that the observed column density is unlikely by the shock heating.

\paragraph{Radiative Excitation by Hot Background Radiation.}
The second candidate is the radiative excitation of cool gas by the hot background radiation in the FIR and (sub)millimeter wavelength, which is a similar process to that of component (b). Thus, we check whether or not reasonable solutions of $T_\mathrm{kin}$, $n_\mathrm{H_2}$, $T_\mathrm{bg}$, and $N_\mathrm{CO}$  exist using RADEX and MCMC, as in Section \ref{subsec:comp_b}. Unlike component (b), the level population of component (c) cannot be observed in the lower rotational levels of $J\le 11$, as shown in Figure \ref{fig:boltz_plot_torus}. Although the gas kinetic temperature cannot be determined as in Section \ref{subsec:comp_b}, we assume that the kinetic temperature is lower or around that of component (b) and set the prior distribution boundary of $T_\mathrm{kin}$ to [2.73, 100] because the temperature should decrease as the rotating radius increases in the torus heated by the central X-ray radiation. The prior distributions of the other parameters and the number and the length of chains are the same as those shown in Section \ref{subsec:comp_b} The integrated autocorrelating steps of each parameter are less than 150; hence, the length is sufficient.\par
\begin{figure}
    \centering
    \gridline{\fig{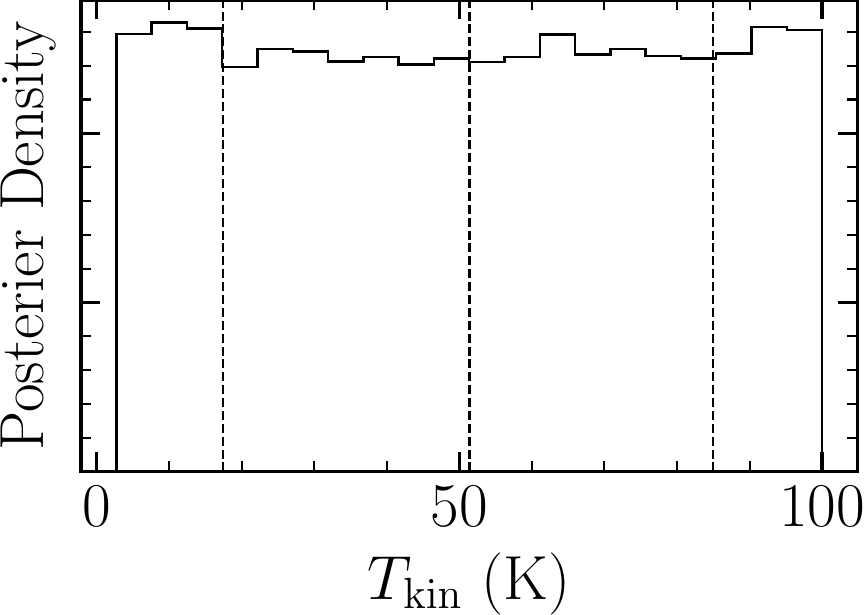}{0.45\linewidth}{}
              \fig{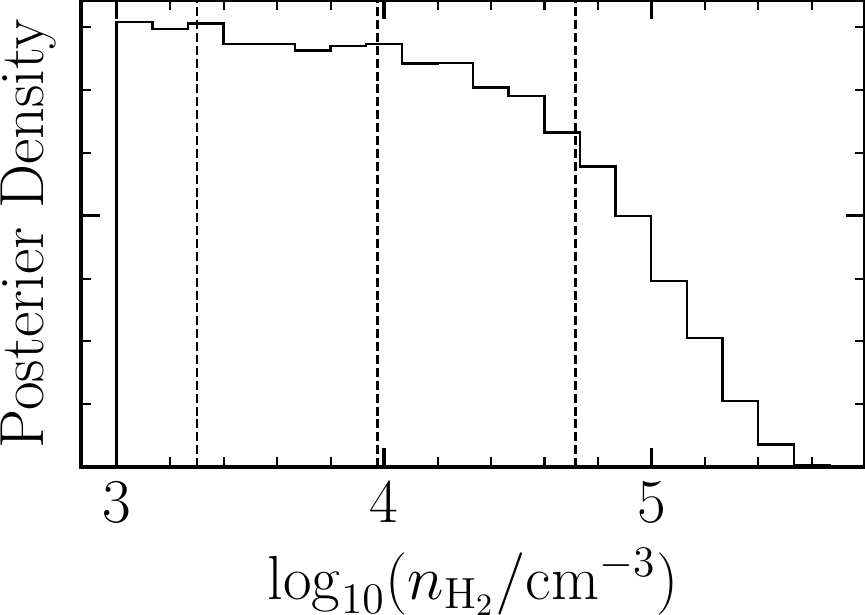}{0.45\linewidth}{}}
    \gridline{\fig{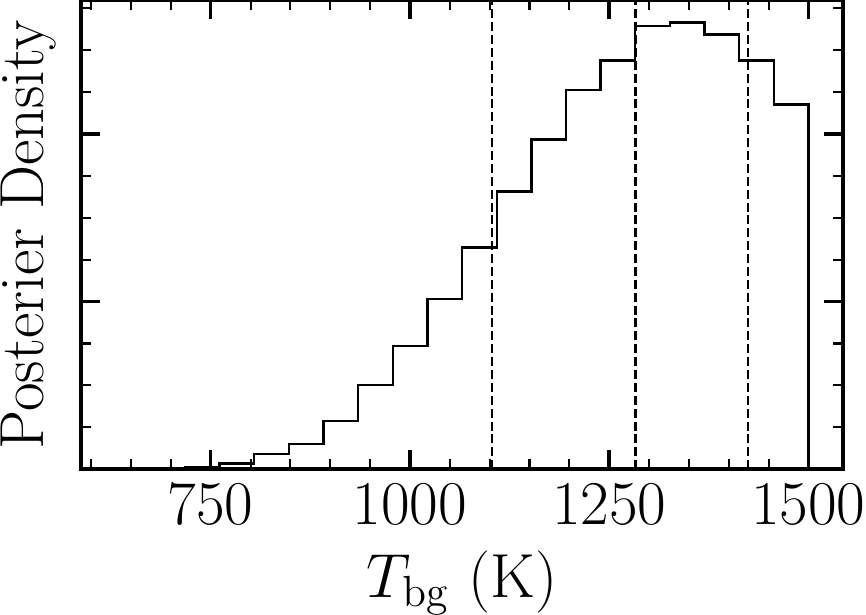}{0.45\linewidth}{}
              \fig{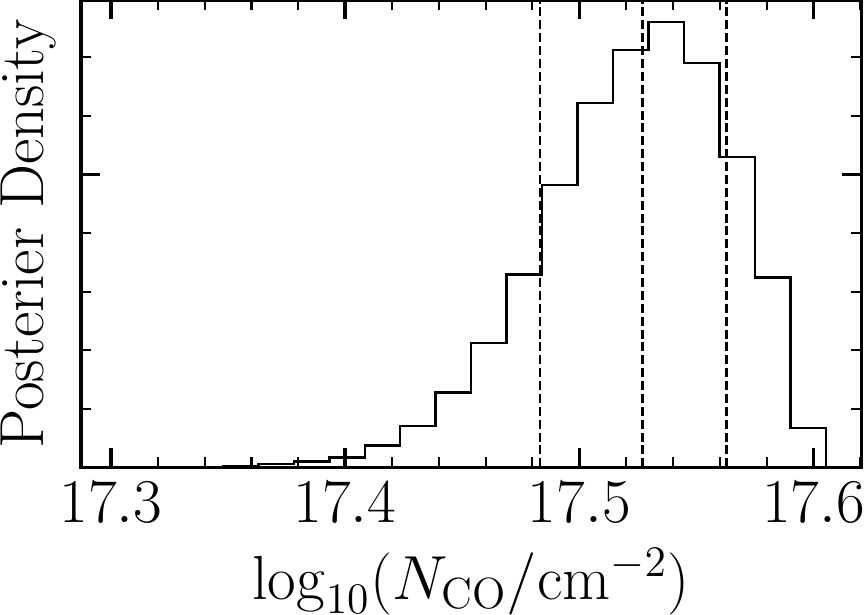}{0.45\linewidth}{}}
    \caption{Posterior distributions of $T_\mathrm{kin}$, $\log\qty(n_\mathrm{H_2}/\mathrm{cm^{-3}})$, $T_\mathrm{bg}$, and $\log\qty(N_\mathrm{CO}/\mathrm{cm^{-2}})$ of component (c). The dashed vertical lines indicate 16th, 50th, and 84th percentiles. The prior boundaries of each parameter are [2.73, 100], [3, 10], [2.73, 1500], and [15, 19]. $T_\mathrm{bg}$ is cut off at $1500\,\mathrm{K}$ owing to the boundary; hence, only the lower limit can be imposed.}\label{fig:mcmchist_comp5}
\end{figure}

We summarize the results from the MCMC in Table \ref{tab:res_mcmc_compbe} and Figure \ref{fig:mcmchist_comp5}. 
Figure \ref{fig:boltz_plot_torus} shows the non-LTE level population of component (c) with the median values of the posterior distributions. The radiative excitation scenario can reproduce the observed level population; thus, we reject the shock excitation scenario and attribute component (c) to the clumps with the density of $\log\qty(n_\mathrm{H_2,c}/\mathrm{cm^{-3}})\le 5.5$ and the CO column density of $\log\qty(N_\mathrm{CO,c}/\mathrm{cm^{-2}})=17.53\pm 0.04$. Here, the upper limit can be imposed on the hydrogen molecular density because dense and cold clumps become LTE and CO molecules cannot be excited to as high rotational levels $J$ as observed. Although we cannot impose any limits to the kinetic temperature, the assumed low kinetic temperature of $T_\mathrm{kin,c}\le 100\,\mathrm{K}$ does not contradict the observed level population. Moreover, only the lower limit can be imposed on the brightness temperature of the FIR-to-(sub)millimeter backgound as $T_\mathrm{bg,c}\ge 784\,\mathrm{K}$.\par
In analogy to components (a) and (b), the lower limit of the geometrical thickness is estimated as
\begin{linenomath}
\begin{align}
\begin{aligned}
    d_\mathrm{los,c}&\sim \frac{N_\mathrm{H_2,c}}{n_\mathrm{H_2,c}}= \frac{[\mathrm{H_2}]}{[\mathrm{CO}]}\frac{N_\mathrm{CO,c}}{n_\mathrm{H_2,c}\phi_V}\\
    &\gtrsim 0.1\left(\frac{\phi_V}{0.03}\right)^{-1}\,\mathrm{pc},
\end{aligned}
\end{align}
\end{linenomath}
where we assume $\mathrm{[CO]/[H_2]}=10^{-4}$ and $\phi_V=0.03$, as in section \ref{subsec:comp_b}. Reasonable solutions exist for the LOS thickness between the lower limit and the typical size of the torus. In conclusion, the origin of component (c) is the inflowing clumps farther from the central black hole than component (b) radiated by the strong FIR-to-(sub)millimeter radiation whose brightness temperature is higher than $784\,\mathrm{K}$.

\section{Discussion}\label{sec:discuss}
In the previous sections, we decomposed the observed CO rovibrational absorption lines into five components (i.e., (a)--(e)) and have found that three components (a)--(c) are attributed to the AGN molecular torus based on the velocity dispersion. The detection of discrete components of the individual LOS velocities or velocity dispersions indicates that the torus medium is not continuous but clumpy.
In this section, we compare the dynamical and physical properties of components (a)--(c) with some torus models in terms of (1) the relation between the LOS velocity and the location, (2) the physical properties, and (3) the temperature gradient.
\subsection{LOS Velocity and Location}\label{subsec:losvel_loc}
\begin{figure}
    \centering
    \includegraphics[width=0.8\linewidth]{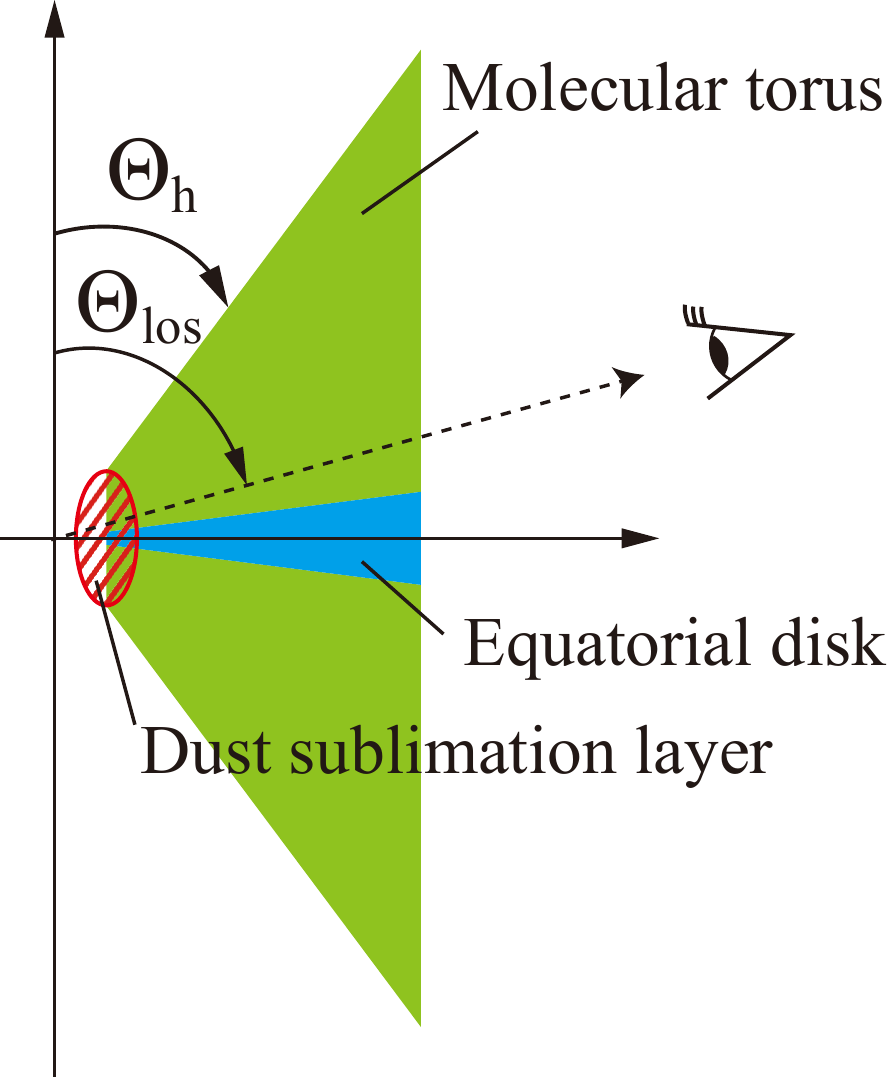}
    \caption{Assumed geometry of the molecular torus and the equatorial disk. The polar angles of the LOS ($\Theta_\mathrm{los}$) and the half-opening angle of the torus ($\Theta_\mathrm{h}$) are also shown.}\label{fig:polar_angles}
\end{figure}

This section discusses whether or not the estimated LOS velocities and locations of components (a)--(c) are consistent with the torus models. The radiation fountain model \citep{Wada2012a,Wada2016} proposes that the outflowing and inflowing gas driven by the radiation from the central accretion disk and the gravity of the central black hole form the molecular torus based on three-dimensional hydrodynamic simulations. \citet{Wada2016} illustrated that the outflowing gas of the velocity of $V_\mathrm{outflow}\lesssim 500\,\mathrm{km\,s^{-1}}$ is naturally reproduced near the center. We compare the outflowing velocity with the observed LOS velocity by assuming the angle between the outflowing direction and the LOS \citet{Vega2008} estimated the half-opening angle of the torus and the LOS angle measured from the pole as $\Theta_\mathrm{h}\sim 40^\circ$ and $\Theta_\mathrm{los}\sim 77^\circ$, respectively, based on the model fitting to the SED of IRAS08 assuming the triangle disk with the constant height-to-radius ratio. Figure \ref{fig:polar_angles} illustrates the assumed geometry of the molecular torus and the polar angles.
Thus, if we assume that the outflowing gas is mainly distributed in the high-latitude region, as suggested in the radiation fountain model, and the outflowing motion is parallel to the boundary between the molecular torus and the ionizing cone, the observable LOS velocity is $V_\mathrm{outflow}\cos(\Theta_\mathrm{los}-\Theta_\mathrm{h})\lesssim 400\,\mathrm{km\,s^{-1}}$, and the observed LOS velocity of the inner outflowing components (a) and (b) of $|V_\mathrm{los}|\sim 160\,\mathrm{km\,s^{-1}}$ is reasonable.\par
On the other hand, the inflowing gas passes through the dense and geometrically thin equatorial disk of the torus \citep{Wada2016,Izumi2018} according to the radiation fountain model. Figure \ref{fig:polar_angles} depicts the equatorial disk. An MHD torus model \citep{Chan2017} also indicates the low-latitude inflow near the equatorial plane in the torus. However, the gas near the equatorial plane is unlikely to be observed as an absorption because the gas column density is too high for the infrared continuum source \citep{Wada2007}. \citet{Wada2016} predicted that the equatorial disk was Compton thick ($2N_\mathrm{H_2}> 1.5\times 10^{24}\,\mathrm{cm^{-2}}$) at $\lesssim 10^\circ$ (or $\Theta_\mathrm{los}\gtrsim 80^\circ$) around the equatorial plane in the Circinus galaxy. Thus, if we assume the physical properties of the equatorial disk to be similar in IRAS08 NW, component (c) is not attributable to such inflowing gas in the thin equatorial disk. Instead, it may be some other inflowing gas, which is apart from the equatorial disk and located in the outer region of the molecular torus, although the origin of the inflowing motion at such a position is unclear at present.\footnote{Although this discrepancy may be caused by the difference between the inner structure of the molecular torus of IRAS08 NW and the Circinus galaxy, we assume the structure to be similar herein.} This scenario is also consistent with the LOS angle of $\Theta_\mathrm{los}\sim 77^\circ\,(\lesssim 80^\circ)$ predicted by \citet{Vega2008}.\par
For the above reasons, the LOS velocity and the location of the inner outflowing components (a) and (b) are consistent with the radiation fountain model. On the other hand, those of the outer inflowing component (c) are not consistent with the inflowing gas passing through the equatorial disk, as suggested in the model, but are likely to be the outer inflowing gas apart from the equatorial disk.

\subsection{Physical Properties}
This section discusses whether or not the kinetic temperature, gas density, and the CO column density of components (a)--(c) are reproduced in theoretical models.\par
In the radiation fountain model \citep{Wada2016}, dense clumps of $n_\mathrm{H_2}\sim 10^6\,\mathrm{cm^{-3}}$ are reproduced, while the gas temperature of the main population with such a high density is lower than $\lesssim 50\,\mathrm{K}$ \citep{Wada2018a}. Hence, component (a), whose density and temperature are $n_\mathrm{H_2}\gtrsim 10^{6.3}\,\mathrm{cm^{-3}}$ and $T_\mathrm{kin,a}\sim 700\,\mathrm{K}$, respectively, is not the main population, while components (b) and (c) are naturally reproduced. However, most of the main populations of such dense gas are unlikely to be observed as the absorption because it is located in the equatorial disk of the torus as mentioned in Section \ref{subsec:losvel_loc}. On the other hand, the CO molecular column density of $N_\mathrm{CO}\sim 10^{18}\,\mathrm{cm^{-2}}$, which is similar to the observed values of $N_\mathrm{CO,a}\sim 10^{18.68}\,\mathrm{cm^{-2}}$ and $N_\mathrm{CO,b}\sim 10^{18.43}\,\mathrm{cm^{-2}}$, is naturally reproduced, even in the regions apart from the equatorial disk \citep{Wada2016}. Thus, it is likely that we selectively observe the rare population that is relatively dense and is located apart from the equatorial plane with the CO rovibrational absorption band.\par
In addition, some clumpy torus models \citep[e.g.,][]{Beckert2004,Vollmer2004,Elitzur2006a} with self-gravitating clumps predict the shear limit for the hydrogen density of each clump to survive against the tidal force of the black hole. For the black hole mass of IRAS08 NW of $M_\mathrm{BH}=9\times 10^7 M_\odot$, the shear limit is given as $n^\mathrm{min}_\mathrm{H}\sim 7\times 10^8\,\mathrm{cm^{-3}}(R/1\,\mathrm{pc})^{-3}$ \citep{Honig2007}. If the rotating radius of component (a) is assumed to be $R_\mathrm{rot,a}\sim 0.5\,\mathrm{pc}$ as in Section \ref{subsec:dist_radii}, the shear limit for components (b) and (c) is $n^\mathrm{min}_\mathrm{H,b}\sim 10^7\,\mathrm{cm^{-3}}$ at $R_\mathrm{rot,b}\sim 2\,\mathrm{pc}$ and $n^\mathrm{min}_\mathrm{H,c}\sim 10^6\,\mathrm{cm^{-3}}$ at $R_\mathrm{rot,c}\sim 8\,\mathrm{pc}$. Thus, the densities of components (b) and (c) are not consistent with the torus models with self-gravitating clumps because the observed densities are less than the shear limits. This indicates that the clumps of components (b) and (c) are not self-gravitating clouds but the density fluctuations as implied in the radiation fountain model \citep{Wada2016} and an MHD torus model \citep{Chan2017}.\par
For the above reasons, the molecular hydrogen density and the kinetic temperature of components (a)--(c) can be reproduced by the radiation fountain model \citep{Wada2018a}, although component (a) is likely to be the rare population selectively observed owing to the equatorial disk obscuration. In addition, components (b) and (c) are unlikely to be self-gravitating as suggested in some clumpy torus models \citep[e.g.,][]{Honig2007} because the hydrogen molecular densities are lower than the shear limits against the central black hole.

\subsection{Temperature Gradient}\label{subsec:temp_grad}
This section discusses the spatial gradient of the gas kinetic temperature between the inner components (a) and (b) to investigate the clumpiness of the torus medium by comparing the gradient with simple XDR models with a uniform gas density and some torus models. Here, we assume that the gas is mainly heated through the photoelectric heating by the local X-ray flux \citep{Maloney1996} and the FIR and (sub)millimeter background discussed in Sections \ref{subsec:comp_b} and \ref{subsec:comp_c} does not significantly affect the kinetic temperature.\par
First, the ratio of the temperatures between components (a) and (b) is derived as
\begin{linenomath}
\begin{gather}
    \frac{T_\mathrm{kin,b}}{T_\mathrm{kin,a}}=\left(\frac{26^{+9}_{-7}}{721\pm 43}\right)=0.036^{+0.013}_{-0.010},
\end{gather}
\end{linenomath}
according to the derived temperatures of components (a) and (b) in Section \ref{subsec:each_comp}.
Second, the ratio of their rotating radii is $R_\mathrm{rot,b}/R_\mathrm{rot,a}=4.8\pm 0.4$ as derived from Equation (\ref{eq:ratio_radii_compab}). Thus, these values lead to the temperature gradient of $T\propto R^{-\alpha}$, whose index is
\begin{linenomath}
\begin{gather}
    \alpha=2.1^{+0.6}_{-0.5},
\end{gather}
\end{linenomath}
if the absorbing medium distributes uniformly between components (a) and (b). The ratio of the rotating radii between components (a) and (b) can be smaller as mentioned in Section \ref{subsec:dist_radii}; thus, this temperature gradient can be steeper. Here, we discuss the gradient of $\alpha\sim 2.1$ as a conservative case where the gradient is mildest.

\paragraph{Comparison with XDR Models.}
We compare this gradient with that predicted in simple XDR models, which assume a uniform gas density distribution (not a clumpy one). In XDR models, the ratio of the local X-ray energy deposition rate per particle ($H_\mathrm{X}/n$) is the controlling parameter \citep{Maloney1996} for the gas temperature; hence, the temperature decreases as the distance from the radiation source increases or the gas density increases. However, we find that such a steep decrease in the temperature from $\sim 720$ to $\sim 25\,\mathrm{K}$ cannot be reproduced, even in the moderately dense medium of the atomic hydrogen density of $n_\mathrm{H}=10^{5.5}\,\mathrm{cm^{-3}}$ if it distributes uniformly according to Figure 3 in \citet{Meijerink2005}.\par
We now discuss whether the temperature gradient can be steeper in a clumpy medium than in a uniform medium. As a simple assumption, we assume that the clumps of components (a) and (b), which have a density of $n_\mathrm{H_2,a}\gtrsim 10^{6.3}\,\mathrm{cm^{-3}}$ and $n_\mathrm{H_2,b}\sim 10^{5.5}\,\mathrm{cm^{-3}}$, are located in the diffuse gas medium. The diffuse gas medium was not observed as CO rovibrational absorption in this work; therefore, the peak optical depth of the absorption by such diffuse gas is likely to be less than the detection limit imposed by the S/N of $\sim 20$. Accordingly, if we assume a line profile function, gas kinetic temperature ($\sim 700\,\mathrm{K}$), and detection limit ($N^\mathrm{min}_{J}\sim 9\times 10^{16}\,\mathrm{cm^{-2}}$ with 3$\sigma$ significance) of the diffuse gas similar to those of component (a), the CO column density of the diffuse gas is less than $N_\mathrm{CO,diffuse}\lesssim 10^{18.2}\,\mathrm{cm^{-2}}$, which is approximately three times smaller than that of component (a) because the maximum $N_{J,\mathrm{a}}$ is approximately three times larger than the detection limit. If this diffuse gas is located uniformly between components (a) and (b), whose separation is approximately $\sim 2\,\mathrm{pc}$, as discussed in Section \ref{subsec:dist_radii}, the 3$\sigma$ upper limit for the hydrogen molecular density of the diffuse gas is $n_\mathrm{H_2,diffuse}\lesssim 10^3\,\mathrm{cm^{-3}}$.\par
For the above reasons, we assume a simple geometry, where components (a) and (b) of $n_\mathrm{H_2,a}\gtrsim 10^{6.3}\,\mathrm{cm^{-3}}$ and $n_\mathrm{H_2,b}\sim 10^{5.5}\,\mathrm{cm^{-3}}$ are located with the ratio of the rotating radii of $R_\mathrm{rot,a}/R_\mathrm{rot,b}\sim 5$ and the diffuse gas of $n_\mathrm{H_2,diffuse}\lesssim 10^3\,\mathrm{cm^{-3}}$ fills the region between them. Figure 3 in \citet{Maloney1996} illustrates that $H_\mathrm{X}/n$ has to decrease by $\sim 3$ orders of magnitude from $\sim 10^{-26}\,\mathrm{erg\,cm^{3}\,s^{-1}}$ at $n=10^3\,\mathrm{cm^{-3}}$ to $\sim 10^{-29}\,\mathrm{erg\,cm^{3}\,s^{-1}}$ at $n=10^5\,\mathrm{cm^{-3}}$ in order for the gas kinetic temperature to decrease from $T\sim 700\,\mathrm{K}$ to $T\sim 25\,\mathrm{K}$. The estimated ratio of the rotating radii between components (a) and (b) is $\sim 5$. The local X-ray energy $H_\mathrm{X}$ then decreases by $\log(5^2)\sim 1$ order of magnitude. In addition, the hydrogen density increases by $\sim 2$ orders of magnitude between the diffuse gas and component (b) clumps. Thus, the apparent temperature gradient is likely to be as steep as $\alpha\sim 2.1$ by the temperature drop due to the decrease of $H_\mathrm{X}/n$ by $\sim 3$ orders of magnitude. Hence, the torus medium has to be clumpy, and the clumps should have the hydrogen atomic density of $n_\mathrm{H}\gtrsim 10^{5.5}\,\mathrm{cm^{-3}}$, which is sufficiently denser than that of the diffuse gas around them.
\paragraph{Comparison with a Clumpy Torus Model.}
The temperature gradient in the torus medium has been estimated by some torus models with more realistic geometries compared to simple XDR models. \citet{Nenkova2008a} estimated the temperature gradient based on their clumpy torus model from the balance between the absorbed energy and the reemitted energy by isolated clumps. If a clump is directly illuminated by the central radiation at the distance from the black hole of $R$, the colder part of the clump has the temperature gradient of $T\propto R^{-0.42}$. This gradient is milder than that derived in this work. Hence, the temperature gradient herein does not support the model where the clumps are directly illuminated by the central radiation, and the extinction of the inner dust should be considered.
\paragraph{Comparison with an MHD Torus Model.}
\citet{Chan2017} also estimated the temperature gradient based on the MHD torus model without assuming any isolated clumps, but attributing the clumps to the density fluctuations in the gas flow. They calculated the temperature gradient based on the balance of the energies emitted and absorbed by a dust grain considering the extinction of the inner dust and derived the temperature gradient of $T\propto R^{-0.73}$. Although this gradient is steeper than that of \citet{Nenkova2008a} owing to the dust extinction, it is also milder than the gradient derived in this work. One of the possible reasons for this is that the clump density is not high enough, as discussed in the comparison with XDR models. Actually, the gas density of the outflowing gas in $|z|/R\gtrsim 0.5$, where $z$ and $R$ are the cylindrical coordinates, fluctuates for one order or less below the hydrogen density of $n_\mathrm{H_2}\sim 10^5\,\mathrm{cm^{-3}}$, according to Figure 12 in \citet{Chan2017}. Here, the AGN luminosity and the black hole mass of IRAS08 NW are assumed to be $L_\mathrm{AGN}\sim 10^{46}\,\mathrm{erg\,s^{-1}}$ \citep{Efstathiou2014} and $M_\mathrm{BH}\sim 9\times 10^7 M_\odot$ \citep{Veilleux2002,Veilleux2009}, respectively.\par
Thus, the temperature gradient derived in this work implies larger density fluctuations than expected in their MHD torus model.

\subsection{Summary of Discussion}
In this section, we have discussed the consistency of the relation between the LOS velocity and the location, the physical properties, and the temperature gradient estimated in this work with some theoretical torus models. First, the radiation fountain model \citep{Wada2016} can naturally explain the LOS velocities, the locations, the hydrogen densities, the kinetic temperatures, and the CO column densities of the outflowing components (a) and (b), although component (a) is not the main population but the rare population selectively observed with the CO rovibrational absorption. On the other hand, another model describing an outer inflow apart from the equatorial disk is necessary to reproduce the inflowing component (c).\par
Second, the clumps of components (b) and (c) are unlikely to be self-gravitating because the hydrogen densities are lower than the shear limits. Thus, they are consistent with the radiation fountain model and MHD models, where the clumps are attributed to the density fluctuations.\par
Third, the temperature gradient between components (a) and (b) is not consistent with simple XDR models with a uniform gas medium, but with a clumpy gas medium. However, an MHD torus model \citep{Chan2017} cannot reproduce the temperature gradient because the density difference caused by the fluctuations is small. Thus, MHD models with larger density fluctuations are necessary to reproduce the temperature gradient.\par
In conclusion, although the radiation fountain model and an MHD model are nearly consistent with the observed results, the larger density fluctuations have to be added to the models in order to reproduce the temperature gradient between components (a) and (b). Moreover, the outer inflows may have to be added to the models to reproduce the location of component (c).

\section{Conclusion} \label{sec:conclusion}
In this paper, we reported the results of the line decomposition of CO rovibrational absorption lines ($v=0\to 1,\ \Delta{J}=\pm 1,\ \lambda\sim 4.67\,\mathrm{\mu{m}}$) in IRAS08 NW to probe the physical parameters and the spatial distribution of each component in each transition. Our findings in this paper are summarized in the following:
\begin{enumerate}
    \item We found five components (a)--(e) in each transition by fitting multiple Gaussians to the optical depth of the observed absorption lines. Among the five components, components (a) and (b) are outflowing with $(V_\mathrm{los}, \sigma_{V})_\mathrm{a}$ $\sim (-160, 175)\,\mathrm{km\,s^{-1}}$ and $(V_\mathrm{los}, \sigma_{V})_\mathrm{b}$ $\sim (-160, 80)\,\mathrm{km\,s^{-1}}$, components (c) and (e) are inflowing with $(V_\mathrm{los}, \sigma_{V})_\mathrm{c}$ $\sim (+100, 42)\,\mathrm{km\,s^{-1}}$ and $(V_\mathrm{los}, \sigma_{V})_\mathrm{e}$ $\sim (+65, 13)\,\mathrm{km\,s^{-1}}$, and component (d) is systemic with $(V_\mathrm{los}, \sigma_{V})_\mathrm{d}$ $\sim (0, 24)\,\mathrm{km\,s^{-1}}$. The LOS velocity is a relative value to the systemic velocity.
    \item The ratios of the rotating radii of components (b)--(e) to component (a) are expected to be ($R_\mathrm{rot,b}/R_\mathrm{rot,a}$, $R_\mathrm{rot,c}/R_\mathrm{rot,a}$, $R_\mathrm{rot,d}/R_\mathrm{rot,a}$, $R_\mathrm{rot,e}/R_\mathrm{rot,a}$) $\approx (5, 17, 53, 180)$ based on the velocity dispersion. Note that these ratios may be the upper limits because the velocity gradient in the molecular torus can be steeper in an MHD model \citep{Chan2017} as mentioned in Section \ref{subsec:dist_radii}. Only components (a)--(c) are attributed to the clumps in the molecular torus because the systemic component (d) and the outer component (e) are attributed to the host galaxy. If we assume that component (a) exists near the dust sublimation layer and $R_\mathrm{rot,a}\approx 0.5\,\mathrm{pc}$, the indicated rotating radii of each component are $R_\mathrm{rot,b}\approx 2\,\mathrm{pc}$ and $R_\mathrm{rot,c}\approx 8\,\mathrm{pc}$ and consistent with the typical torus scales. These ratios indicate the velocity structure of the torus, where the clumps in the inner regions are outflowing while those in the outer regions are inflowing, as shown in Figure \ref{fig:torus_geom}.
    \item Based on the level population, component (a) is attributed to the hot ($T_\mathrm{kin,a}\sim 700\,\mathrm{K}$) and dense ($n_\mathrm{H_2,a}\gtrsim 10^{6.3}\,\mathrm{cm^{-3}}$) clumps that are in the LTE and located innermost of components (a)--(c). On the other hand, the outer components (b) and (c) are attributed to the non-LTE clumps radiatively excited by the FIR-to-(sub)millimeter background radiation fields, whose brightness temperatures are $T_\mathrm{bg,b}\ge 236\,\mathrm{K}$ and $T_\mathrm{bg,c}\ge 784\,\mathrm{K}$, respectively, whereas they are likely to have a low kinetic temperature of $T_\mathrm{kin}\sim 25\,\mathrm{K}$.
    \item The LOS velocities and the locations of outflowing components (a) and (b) are consistent with the radiation fountain model \citep{Wada2012a,Wada2016}. On the other hand, those of inflowing component (c) are not attributable to the low-latitude inflow predicted in the model and MHD models \citep{Chan2017}, but to some other inflowing gas located in the outer region of the molecular torus.
    \item The radiation fountain model naturally reproduces the kinetic temperature and the gas density of components (b) and (c), while component (a) shows a much higher temperature than the main population with the density of $n_\mathrm{H_2}\sim 10^6\ \mathrm{cm^{-3}}$ in the model. This is because we selectively observe the dense clumps apart from the dense equatorial plane of the torus owing to the obscuration of the infrared continuum source by the plane \citep{Wada2007}.
    \item We discussed the radial gradient of the kinetic temperature between components (a) and (b) based on the assumption that the gas was predominantly heated by the local X-ray flux photoelectrically. According to XDR models, the gas kinetic temperature gradient between outflowing components (a) and (b) of $\alpha\sim 2.1$ in $T_\mathrm{kin}\propto R_\mathrm{rot}^{-\alpha}$ agrees with the scenario that the torus has a clumpy medium.
    \item Although the radiation fountain model \citep{Wada2012a,Wada2016} and an MHD model \citep{Chan2017} are almost consistent with the observed results, the outer inflows and the larger density fluctuations may have to be added to the models in order to reproduce the location of component (c) and the temperature gradient between components (a) and (b).
\end{enumerate}

\begin{acknowledgments}
    The authors thank the anonymous referee for many comments and suggestions to improve this paper. We also thank all the operating staff in Subaru Telescope and NAOJ for a great help in our observation. Especially, we are grateful to Dr. Tae-Soo Pyo, Dr. Takagi, and Dr. Mieda for a lot of supports in the IRCS observations in 2010 and 2019. In addition, we thank Mr. Doi for many fruitful discussions. This work was supported by JSPS KAKENHI grant Nos. JP19J21010 (S.O.) and JP19J00892 (S.B.).
\end{acknowledgments}

%

\facility{Subaru (IRCS)}


\software{IRAF v2.16.1 \citep{Tody1986,Tody1993}, PyRAF v2.1.15 \citep{Pyraf}, Molecfit v1.5.9 \citep{Kausch2015,Smette2015}, TIPS \citep{Gamache2017}, RADEX v08sep2017 \citep{vanderTak2007}, Numpy v1.18.5 \citep{Harris2020}, Matplotlib v3.2.2 \citep{Hunter2007a}, Scipy v1.4.1 \citep{Virtanen2020}, Astropy v4.0.1 \citep{Astropy2013,Astropy2018}, Lmfit v1.0.0 \citep{Lmfit}, Emcee v3.0.2 \citep{Foreman-Mackey2013}, Pandas v1.0.3 \citep{McKinney2010,Pandas}, Jupyter v1.0.0 \citep{Kluyver2016}, ASTEVAL v0.9.18 \citep{Asteval}}

\appendix
\section{Effects of Area Covering Factors}\label{app:area_cov}
We evaluate the effect of an area covering factor of each component, which is assumed to be unity in this paper, on the parameter estimations in Section \ref{subsec:each_comp}. Here, we concentrate on the effects on the CO column density and the kinetic temperature because they are important parameters in the discussions in Section \ref{sec:discuss}.\par
If clumps have the area covering factor of $C_f$, the optical depth $\tau(\lambda)$ and the normalized flux $F_\lambda/F_\mathrm{c}$ are related as
\begin{linenomath}
    \begin{gather}
        \tau(\lambda)=-\ln\left[\frac{C_f-(1-F_\lambda/F_\mathrm{c})}{C_f}\right]\propto N_J.
    \end{gather}
\end{linenomath}
In the fully covered case with $C_f=1$, this equation is identical to that given in Section \ref{subsec:analysis_model_fit}. Accordingly, the CO column density at ($v=0,\ J$), which is $N_J$, becomes larger than that estimated based on the fully covered assumption with $C_f=1$ as the $(1-F_\lambda/F_\mathrm{c})$ values at the absorption peaks of each component approach $C_f$ or $C_f$ itself becomes smaller.\par
At the extreme case where the area covering factors are the smallest, those of components (a)--(c) are $C^{\mathrm{min}}_{f,\mathrm{a}}= 0.356$, $C^{\mathrm{min}}_{f,\mathrm{b}}= 0.635$, and $C^{\mathrm{min}}_{f,\mathrm{c}}= 0.157$ based on the $(1-F_\lambda/F_\mathrm{c})$ values at the absorption peaks of each component in $R(13)$, $R(3)$, and $R(12)$ in which each component gives the deepest absorption. The CO column density of component (c) can differ the most of the three components because the covering factor is the smallest. The CO molecular column density ($N_\mathrm{CO}$) is mainly determined by the level population at lower rotational levels; thus, we evaluate the difference of $N_{J,\mathrm{c}}$ at $J=6$. The CO column density at $J=6$ with $C_{f,\mathrm{c}}=C^{\mathrm{min}}_{f,\mathrm{c}}= 0.157$ then becomes approximately seven times larger than that with $C_{f,\mathrm{c}}= 1$ because $(1-F_\lambda/F_\mathrm{c})_{J=6,\mathrm{c}}\sim 0.037$ at the absorption peak (Figure \ref{fig:fitres_colines}). In short, the CO molecular column density ($N_\mathrm{CO}$) can be at most approximately seven times larger than that based on the fully covered assumption because of the covering factor correction.\par
On the other hand, kinetic temperature is determined only at components (a) and (b). Hence, we evaluate the effects of the covering factor in component (a) with the smaller $C^{\mathrm{min}}_f$. The excitation temperature $T_{\mathrm{ex},JJ'}$ based on the $N_J/N_{J'}$ $(J'>J)$ is written as
\begin{linenomath}
    \begin{multline}
        T_{\mathrm{ex},JJ'}(C_f=1)\\
        =\frac{E_{J'}-E_J}{k_\mathrm{B}}\left[\ln\left(\frac{N_J}{N_{J'}}\cdot\frac{2J'+1}{2J+1}\right)\right]^{-1},
    \end{multline}
\end{linenomath}
in the fully covered case. Then, if the column densities $N_J$ and $N_{J'}$ are magnified by a factor of $m_J$ and $m_{J'}$, respectively, owing to the covering factor correction, the estimated excitation temperature differs as
\begin{linenomath}
    \begin{gather}
        T_{\mathrm{ex},JJ'}(C_f)=T_{\mathrm{ex},JJ'}(C_f=1)+\Delta{T_{\mathrm{ex},JJ'}},\\
        \Delta{T_{\mathrm{ex},JJ'}}\equiv -\frac{E_{J'}-E_J}{k_\mathrm{B}}\ln\left(\frac{m_J}{m_{J'}}\right),
    \end{gather}
\end{linenomath}
when $T_{\mathrm{ex},JJ'}(C_f=1)\gg \Delta{T_{\mathrm{ex},JJ'}}$. The kinetic temperature is mainly determined by the excitation temperature based on the level population at lower rotational levels; hence, we evaluate $\Delta{T_{\mathrm{ex},JJ'}}$ based on the ratio of $N_{J=3,\mathrm{a}}/N_{J=12,\mathrm{a}}$. If the covering factor is $C_{f,\mathrm{a}}=C^{\mathrm{min}}_{f,\mathrm{a}}=0.356$, the magnifications caused by the covering factor are $m_{J=3}\sim 3.4$ and $m_{J=12}\sim 5.3$ based on the absorption peaks of $(1-F_\lambda/F_\mathrm{c})_{J=3,\mathrm{a}}\sim 0.15$ and $(1-F_\lambda/F_\mathrm{c})_{J=12,\mathrm{a}}\sim 0.30$. The excitation temperature then increases by $\Delta{T_{\mathrm{ex},JJ'}}\sim 180\,\mathrm{K}$, which is $\sim 25\%$ of $T_\mathrm{ex,a}(C_f=1)\sim 720\,\mathrm{K}$.\par
In conclusion, the covering factors do not greatly change the conclusions in this work because the parameters are not very different under the conditions discussed above.

\section{Ice Features}\label{sec:ice_feature}
This appendix summarizes the results of the detected ice features in the process of the model fitting to the gaseous CO absorption, as mentioned in Section \ref{subsec:analysis_model_fit}.\par
\begin{figure}
    \centering
    \gridline{\fig{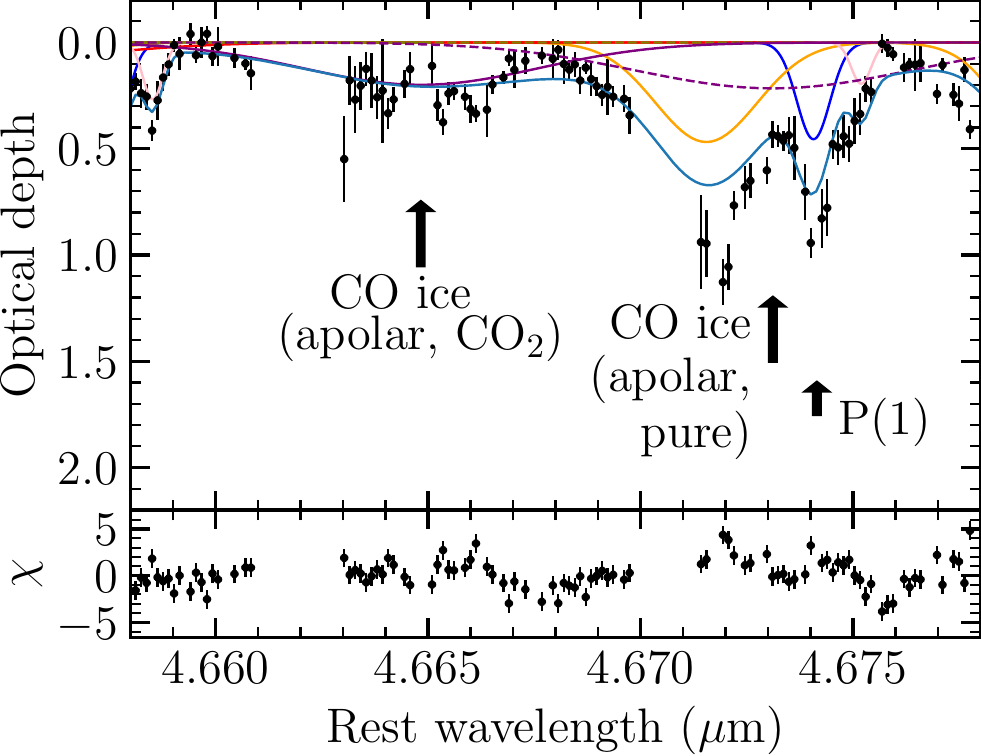}{0.95\linewidth}{}}
    \gridline{\fig{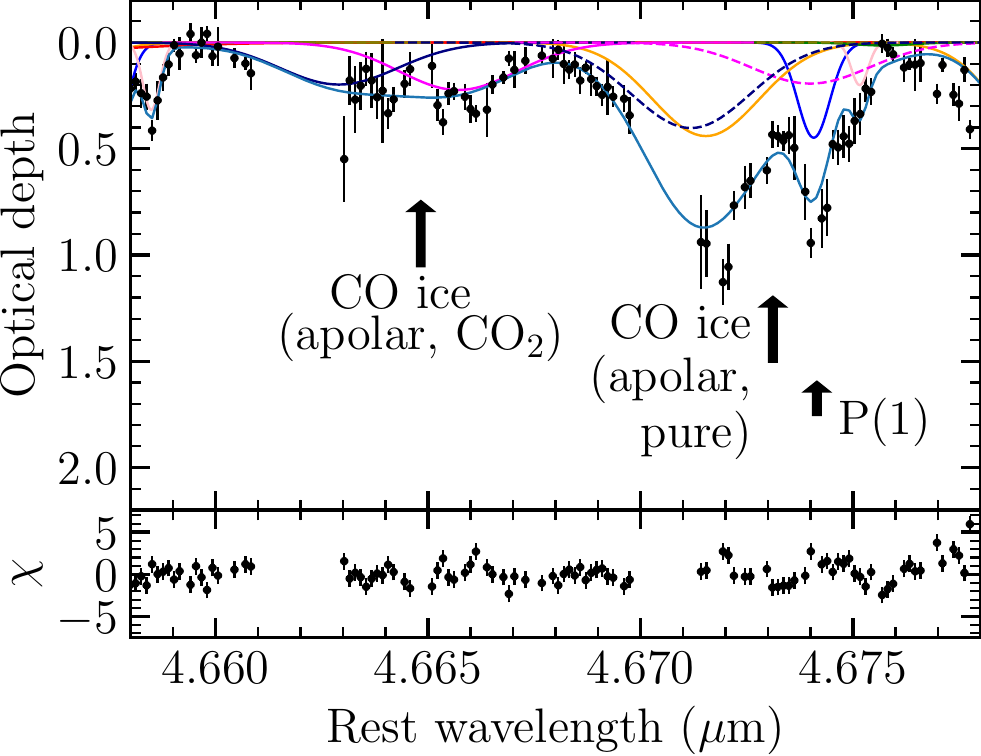}{0.95\linewidth}{}}
    \caption{All the notations are the same as those in Figure \ref{fig:fitres_colines}. Top: the best-fit model when each CO ice feature consists of one Gaussian component at the upper half, and the residual significance at the lower half. Bottom: the best-fit model when each CO ice feature consists of two Gaussian components which are blueshifted and redshifted at the upper half, and the residual significance at the lower half.}
    \label{fig:coice_fit}
\end{figure}

Figure \ref{fig:coice_fit} shows the detected gaseous CO and ice features around the band center. Based on the central wavelength of the features, we attributed the feature around $\lambda\sim 4.665\,\mathrm{\mu{m}}$ to the $\mathrm{CO_2}$-dominant apolar CO ice and that around $\lambda\sim 4.673\,\mathrm{\mu{m}}$ to the pure apolar CO ice. In addition, we introduced apolar and polar $\mathrm{OCN^-}$ ice at $\lambda\sim 4.60\,\mathrm{\mu{m}}$ as shown in Figure \ref{fig:fitres_colines}. According to their mixture partner, they exist as apolar ice dominated by molecules with low dipole moments, or polar ice dominated by molecules with high dipole moments, and their absorption features are centered at different wavelengths.\par
The central wavelength and the FWHM band width of the $\mathrm{CO_2}$-dominant apolar CO ice and pure apolar CO ice are $(\lambda_\mathrm{ice},\Delta{\lambda})=(4.665\,\mathrm{\mu{m}},0.0065\,\mathrm{\mu{m}})$ and $(\lambda_\mathrm{ice},\Delta{\lambda})=(4.673\,\mathrm{\mu{m}},0.0076\,\mathrm{\mu{m}})$, respectively \citep{Pontoppidan2003,Boogert2015b}. Meanwhile, the central wavelength and the FWHM band width of the apolar and polar $\mathrm{OCN^-}$ ice are $(\lambda_\mathrm{ice},\Delta{\lambda})=(4.598\,\mathrm{\mu{m}},0.032\,\mathrm{\mu{m}})$ and $(\lambda_\mathrm{ice},\Delta{\lambda})=(4.617\,\mathrm{\mu{m}},0.055\,\mathrm{\mu{m}})$, respectively \citep{vanBroekhuizen2005,Boogert2015b}. These ice features were detected for the first time in the extragalactic environment in a starburst NGC 4945 by \citet{Spoon2000,Spoon2003a}, whereas they had been detected in many protostellar objects. Although \citet{Spoon2003a} attributed the ice features at $\lambda\sim 4.60\,\mathrm{\mu{m}}$ to ``XCN'' ice, which is defined as ice including C$\equiv$N bondings, we attribute them herein to the molecular ion $\mathrm{OCN^-}$ as in \citet{Demyk1998a}.\par
To estimate the column density of each ice feature ($N_\mathrm{ice}$), we fitted the Gaussian to the optical depth ($\tau_\lambda$) as follows:
\begin{linenomath}
    \begin{gather}
        \tau_\lambda=\frac{\lambda_\mathrm{ice}^2A_\mathrm{ice}N_\mathrm{ice}}{\sqrt{2\pi}\sigma_\lambda}\exp\left[-\frac{(\lambda-\lambda_\mathrm{ice})^2}{2\sigma_\lambda^2}\right],\\
        \sigma_\lambda=\frac{\Delta\lambda}{2\sqrt{2\ln 2}}
    \end{gather}
\end{linenomath}
where $A_\mathrm{ice}$ is the integrated band strength of the ice feature. The integrated band strengths of CO ice and $\mathrm{OCN^-}$ ice are assumed to be $1.1\times 10^{-17}\,\mathrm{cm\,molecule^{-1}}$ \citep{Gerakines1995} and $1.3\times 10^{-16}\,\mathrm{cm\,molecule^{-1}}$ \citep{vanBroekhuizen2004}, respectively.\par
Figure \ref{fig:coice_fit} shows the best-fit results of CO ice.
The observed CO ice absorption bands were double-peaked and narrower than those previously observed in the protostellar objects \citep{Lacy1984a,Pontoppidan2003}. As shown in the top panel of Figure \ref{fig:coice_fit}, single Gaussians with previously observed widths (FWHMs of the $\mathrm{CO_2}$-mixed CO ice and pure CO ice are $0.0065\,\mathrm{\mu{m}}$ and $0.0076\,\mathrm{\mu{m}}$, respectively.) result in an excess around the peak at $\lambda\sim 4.672\,\mathrm{\mu{m}}$ and a deficit around the valley at $\lambda\sim 4.676\,\mathrm{\mu{m}}$ over 3$\sigma$ significance, indicating that the fitted models have too broad widths. According to the laboratory measurements by \citet{vanBroekhuizen2006}, the FWHMs of pure CO ice and layered $\mathrm{CO_2/CO}$ ice are $\Delta\lambda=0.0033\,\mathrm{\mu{m}} (\approx 1.5\,\mathrm{cm^{-1}})$, which is the half as broad as those previously observed, at the ice temperature of $\lesssim 25\,\mathrm{K}$. Thus, we have fitted each CO ice feature with two narrower Gaussians with these fixed FWHMs as illustrated in the bottom panel of Figure \ref{fig:coice_fit}. For the first estimate, we have tied each central velocity of the two gaussians between pure CO ice and $\mathrm{CO_2}$-mixed CO ice, and the estimated velocity shifts are $-123\pm 14\,\mathrm{km\,s^{-1}}$ and $56\pm 19\,\mathrm{km\,s^{-1}}$. It is unclear whether these shifts are the real velocity shifts or the shape irregularity of CO ice absorption bands such as ``core--mantle effects'' suggested by \citet{Tielens1991}.\par
Table \ref{tab:fitres_ice} summarizes the estimated column density of each ice. The column density of each CO ice is the sum of the redshifted and blueshifted components. The thermal history of the ice features and the origins of the redshifted and blueshifted components in CO ice will be discussed in the future work.

\begin{deluxetable}{lcc}
\tablecaption{Estimated Column Densities of the Ice Features
\label{tab:fitres_ice}}
\tablehead{
\colhead{Ice} & \colhead{$\lambda_\mathrm{ice}\ (\mathrm{\mu{m}})$} & \colhead{$N\ \mathrm{(10^{16}\ cm^{-2})}$}
}
\decimalcolnumbers
\startdata
Apolar $\mathrm{OCN^-}$ & 4.598 & 2.2 $\pm$ 0.4\\
Polar $\mathrm{OCN^-}$ & 4.617 & $\le$ 1.2\\
Apolar CO ($\mathrm{CO_2}$ dominant) & 4.665 & 6 $\pm$ 1\\
Apolar CO (pure) & 4.673 & 9 $\pm$ 1
\enddata
\tablecomments{
Column (2): the central wavelength. References are \citet{vanBroekhuizen2005} for $\mathrm{OCN^-}$ ice and \cite{Pontoppidan2003} for CO ice. Column (3): the estimated column density. The column density of each CO ice is the sum of the redshifted and blueshifted components. (See the text for the details.)
}
\end{deluxetable}

\section{Temperature and Column Density of the Outer Components}\label{app:host_gal}
This appendix summarizes the excitation temperature and the column density of components (d) and (e), which are expected to be located in the outer regions. The parameters are estimated by fitting a Boltzmann distribution to the level population as performed for component (a) in Section \ref{subsec:comp_a}.
\begin{figure}
    \centering
    \includegraphics[width=0.95\linewidth]{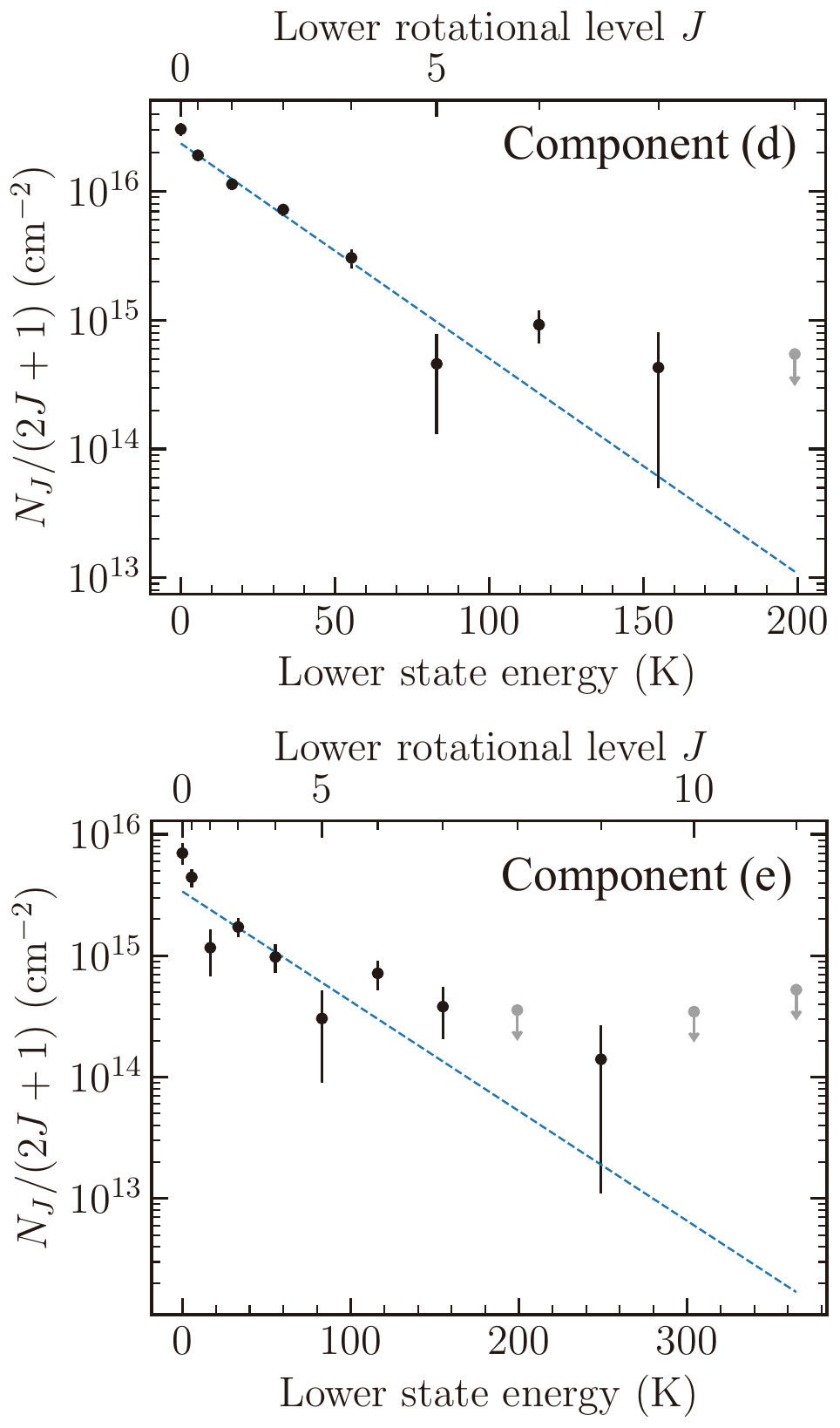}
    \caption{Population diagrams of components (d) and (e). The notations are the same as those in Figure \ref{fig:boltz_plot_torus}.}
    \label{fig:boltz_plot_others}
\end{figure}
Figure \ref{fig:boltz_plot_others} illustrates the population diagrams of components (d) and (e).\par
As for component (d), the excitation temperature and the CO column density are $T_\mathrm{ex,d}=26\pm 3\,\mathrm{K}$ and $N_\mathrm{CO,d}=(2.3\pm 0.2)\times 10^{17}\,\mathrm{cm^{-2}}$, respectively. The low excitation temperature of $T_\mathrm{ex}\sim 26\,\mathrm{K}$ is consistent with the assumption that this component is attributed to the host galaxy, as mentioned in Section \ref{subsec:dist_radii}.\par
Similarly, as for component (e), the excitation temperature and the CO column density are $T_\mathrm{ex,e}=48\pm 16\,\mathrm{K}$ and $N_\mathrm{CO,e}=(6\pm 1)\times 10^{16}\,\mathrm{cm^{-2}}$, respectively. Component (e) is so narrow that it is unresolvable, and the bulk motion is an infall. This component is not resolved, and the exact dynamical states are unclear; hence, we do not discuss its origin.


\bibliographystyle{aasjournal}
\bibliography{library_ext}



\end{document}